\documentclass[letter, 10pt, twocolumn]{article}

\usepackage{amsmath}
\usepackage{amsthm}
\usepackage{amssymb}
\usepackage{mathtools}

\usepackage[table, dvipsnames]{xcolor}
\usepackage[shortlabels]{enumitem}
\usepackage{booktabs}
\usepackage{chemformula}
\usepackage[ruled]{algorithm2e}
\usepackage[colorlinks=true, linkcolor=black, citecolor=ForestGreen, urlcolor=cyan]{hyperref}
\usepackage{caption}
\usepackage{subcaption}
\usepackage[capitalise, noabbrev]{cleveref}
\usepackage{authblk}
\usepackage{doi}

\crefformat{footnote}{#2\footnotemark[#1]#3}

\usepackage[symbol]{footmisc}

\usepackage[UKenglish]{babel}
\usepackage{csquotes}
\usepackage[shortcuts]{extdash}
\usepackage[useregional]{datetime2}

\usepackage{graphicx}

\usepackage{parskip}
\usepackage{microtype}
\usepackage[left=1.2cm, right=1.2cm]{geometry}

\usepackage[notextcomp]{stix2}
\usepackage[font=small]{caption}
\usepackage[super,comma,sort&compress]{natbib}

\title{Trading off regional and overall energy system design flexibility in the net-zero transition}

\author[1]{Koen van Greevenbroek\footnote{Contributed equally}\thanks{Corresponding author, \url{koen.v.greevenbroek@uit.no}}}
\author[2,3]{Aleksander Grochowicz\protect\footnotemark[1]}
\author[4]{Marianne Zeyringer}
\author[2]{Fred Espen Benth}

\affil[1]{Department of Computer Science, UiT The Arctic University of Norway, Postboks 6050 Langnes, 9037 Tromsø, Norway}
\affil[2]{Department of Mathematics, University of Oslo, P.O. Box 1053 Blindern, 0316 Oslo, Norway}
\affil[3]{Department of Wind and Energy Systems, Technical University of Denmark, Elektrovej 325, 2800 Kgs. Lyngby, Denmark}
\affil[4]{Department of Technology Systems, University of Oslo, P.O. Box 70, 2027 Kjeller, Norway}

\date{\today}

\begin{document}

\twocolumn[%

\maketitle

% \bfseries%
%TC:ignore

\begin{center}
    \begin{minipage}{0.75\textwidth}
        \subsection*{Abstract}
        The transition to net-zero emissions in Europe is determined by a patchwork of country-level and EU-wide policy, creating coordination challenges in an interconnected system.
        We use an optimisation model to map out near-optimal energy system designs for 2050, focussing on the planning flexibility of individual regions while maintaining overall system robustness against different weather years, cost assumptions, and land use limitations.
        Our results reveal extensive flexibility at a regional level, where only few technologies (solar around the Adriatic and wind on the British Isles and in Germany) cannot be substituted.
        National policymakers can influence renewable energy export and hydrogen strategies significantly, provided they coordinate this with the remaining European system.
        However, stronger commitment to solar in Southern Europe and Germany unlocks more design options for Europe overall.
        These results on regional trade-offs facilitate more meaningful policy discussions which are crucial in the transition to a sustainable energy system.
    \end{minipage}
\end{center}

%TC:endignore

\vspace{3ex}
]

\footnotetext[1]{Contributed equally}
\footnotetext[2]{Corresponding author, \url{koen.v.greevenbroek@uit.no}}

The target of net carbon neutrality by 2050 laid out in the European Climate Law \cite{eu-climate-law-2021} will lead to a massive transformation of the European energy system both at the domestic and international level.
The EU plans to achieve the transition through a mix of the European Green Deal \cite{eu-green-deal-2019} and national policy with a focus on inclusion, justice, and sustainability.
However, the energy transition is often modelled with a central planner approach \cite{hofbauer-mcdowall-ea-2022,shu-reinert-ea-2024}, disregarding national interests and diverse issues such as social acceptance, energy prices and security.
Though the regional dynamics of decision making are crucial for investment planning \cite{kendziorski-goke-ea-2022,durakovic-delgranado-ea-2023,strambo-nilsson-ea-2015}, systematic studies delineating the renewable investment options of individual regions embedded in a larger system illuminating trade-offs and interactions are missing.

Near-optimal spaces of energy system models \cite{decarolis-2011,neumann-brown-2021,pedersen-victoria-ea-2021,grochowicz-vangreevenbroek-ea-2023} effectively generate different alternatives for the energy transition and these techniques have demonstrated a large degree of design flexibility in a net-zero European energy system by 2050 \cite{neumann-brown-2023,pickering-lombardi-ea-2022,grochowicz-vangreevenbroek-ea-2023}.
Allowing for a slight cost increase can improve system resilience \cite{grochowicz-vangreevenbroek-ea-2023} and make space for exploring qualitative and socio-political aspects \cite{vagero-jacksoninderberg-ea-2024} that are under the radar in a cost optimisation.
Some studies have investigated spatially diverse energy system designs for Europe \cite{pickering-lombardi-ea-2022,pedersen-andersen-ea-2023,sasse-trutnevyte-2023a}, but have not systematically mapped out the options for renewable investment in individual regions within Europe.

To address the research gap of analysing regional trade-offs, we use near-optimal methods to conduct a rigorous study of the effects of regional decision-making on the future European energy system and vice versa.
We build on the PyPSA-Eur-Sec model \cite{PyPSAEurSec,neumann-zeyen-ea-2023a} to conduct partial greenfield optimisations for a net-zero emissions system including the electricity, heating, transportation and industry sectors with a 2050 planning horizon (Methods).
Using a diverse set of scenarios to ensure robustness to uncertainty, we obtain results for 7 similarly sized \emph{focus regions} embedded in the European energy system.
This enables a comparison of regional design flexibility, export potentials and minimum investment levels subject to various system-wide investment decisions.

We find a strikingly large degree of design flexibility --- on the continental, but even more so on the regional level.
If renewable generation is sufficient, policymakers may trade off different objectives among many cost-effective alternatives.
This allows the prioritisation of natural resources, economic interests, social acceptance or other factors while ensuring a joint and near-optimal transition on the European scale.
Still, investment in certain technologies and regions does affect overall system design flexibility much more than others, with for example solar power in Germany providing flexibility both domestically and system-wide.
Wind and hydrogen investment, meanwhile, can be geographically shifted relatively freely, revealing investment opportunities for individual regions to become renewable energy exporters or hydrogen powerhouses.
This significantly expands the understanding of design flexibility within the renewable electricity \cite{grochowicz-vangreevenbroek-ea-2023,neumann-brown-2023} or energy sector \cite{pickering-lombardi-ea-2022} in Europe, precisely describing the considerable trade-offs between solar, onshore- and offshore wind at a regional resolution.

Our innovation is to map out near-optimal spaces in a set of dimensions representing investment in key technologies both \emph{inside} and \emph{outside} a given region.
We concentrate on solar, onshore wind, offshore wind and hydrogen infrastructure (production, storage, conversion, transportation), resulting in a total of 8 dimensions (4 inside, 4 outside).
Approximations of the resulting joint near-optimal design spaces \cite{grochowicz-vangreevenbroek-ea-2023} are then computed for each of the 7 focus regions separately.
Crucially, we ensure robustness of the results against important uncertainties by intersecting near-optimal spaces arising from 12 different scenarios.
These include a baseline scenario, scenarios with higher costs for solar and wind and one with more restricted land availability for solar; each of these is run with 3 different weather years (Methods).
For the remainder of this work, we refer to the intersection of near-optimal spaces as the \emph{robust design space}, and to points therein as \emph{robust solutions/designs}.
The intersection technique has previously only been applied to weather years \cite{grochowicz-vangreevenbroek-ea-2023}; our approach involving a variety of scenarios provides a novel blueprint for future studies of robust system design.
The present study is also the first to systematically map out the European robust design space for solar, onshore \& offshore wind and hydrogen infrastructure dimensions jointly in a sector-coupled context.

\section*{Results}
\subsection*{Ample opportunities for regional energy supply and exports}

\begin{figure*}
    \centering
    \includegraphics[width=18cm]{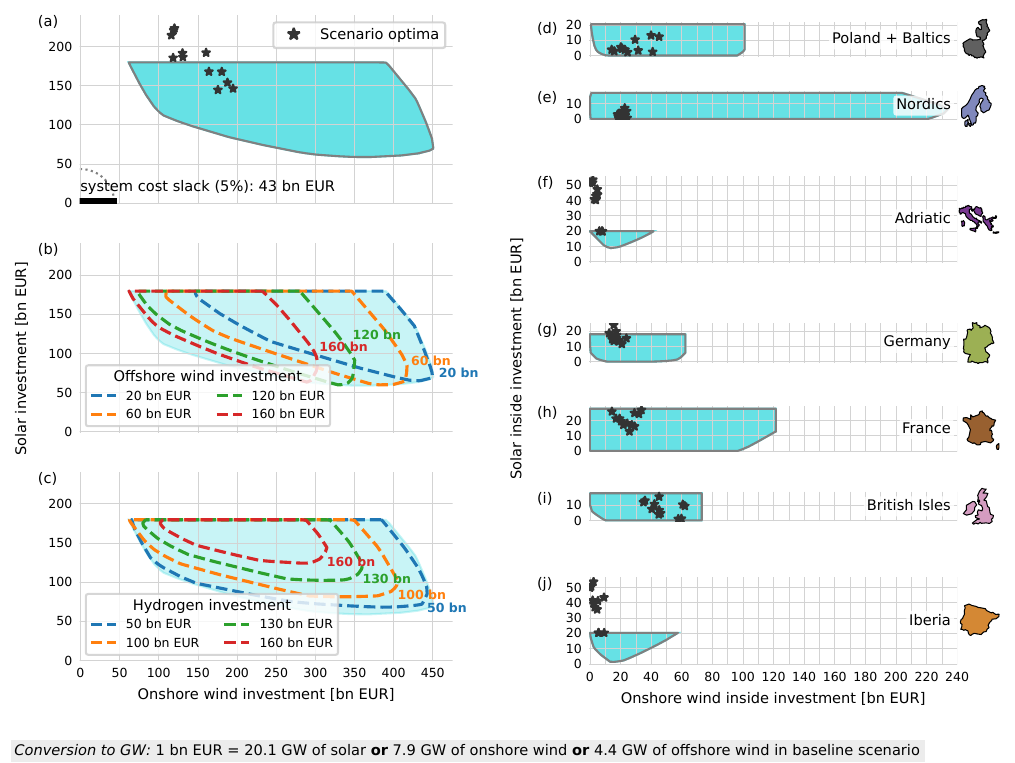}
    \caption{Robust trade-offs between annualised onshore wind and solar investment in Europe, consistent with the net-zero emissions target for 2050. (a) shows the overall robust design space projected onto the onshore wind and solar dimensions, with the cost slack marked, as well as the optima in all the individual scenarios marked. Note that the optima from individual scenarios are not necessarily robust (i.e.\ may not be feasible or near-optimal in other scenarios). (b) -- (c) show how the robust design spaces changes (in dashed lines) subject to various levels of investment in offshore wind and hydrogen infrastructure respectively. For instance, investing only 20 bn EUR into offshore wind requires more investment in solar and/or onshore wind.
    Note that as cost-optimal European-wide hydrogen infrastructure investment is on average 67 bn EUR, investment beyond this level takes away cost slack from other technologies (here, solar and onshore wind) and forces their values closer to optimum, reducing the extent of the design space.
    (d) -- (j) show the trade-offs between investment in solar PV and onshore wind for the different focus regions and display the vast opportunities of robust and cost-effective onshore wind-solar substitution.
    Solar investment is limited in robust designs by the scenarios allowing only a maximum average installation density of 1.7~MW/km$^2$ for utility solar within available land (corresponding to roughly 1.2\% of the total land area of the modelling region); see Supplementary Figures S3--5 for alternative versions of this figure with a looser land-use restriction for solar and including land-use restrictions for wind power.
    The conversion guide from bn EUR (annualised) to GW for different technologies, shown at the bottom, is valid for all figures in the baseline scenarios. In scenarios where solar or wind power are more expensive, the same investment will result in smaller capacities.}
    \label{fig:overview-onwind-solar}
\end{figure*}

On both system-wide and regional levels, there is significant flexibility in how investment can be distributed onto renewable technologies and hydrogen infrastructure.
This flexibility holds even while taking into account uncertainties in costs, weather years, and land use restrictions, extending previous results using near-optimal methods \cite{pickering-lombardi-ea-2022,neumann-brown-2023}.
Indeed, we only consider so-called robust solutions that are feasible and cost-effective under all considered scenarios (Methods).
Our results are given for a total system cost slack of $\varepsilon = 5\%$, but see Supplementary Note S1, Table S1 and Figures S1 for a sensitivity analysis of key figures under different slack levels.
Relatively speaking, individual regions enjoy significantly more planning flexibility than we see for the overall system (\cref{fig:overview-onwind-solar}).

We see that any robust European system design needs at least around 60 bn EUR investment in onshore wind and solar power each, amounting to $\sim$500~GW and $\sim$1250~GW of installed onshore wind and solar capacity, versus 188~GW \cite{windeurope-2023} and 209~GW \cite{michaelschmela-2022} in EU-27 in 2022, respectively. 
This combination of wind (500~GW) and solar power (1250~GW) is not enough; remaining investment (of at least 340 bn EUR continent-wide) can be distributed in many different ways (\cref{fig:overview-onwind-solar} (a)--(c)).
Details on technologies beyond wind and solar are given in Supplementary Figure S2.
All costs are annualised with a 7\% discount rate and given in 2023 EUR.

We model for net zero \ch{CO2} emissions and limit \ch{CO2} sequestration to 200 Mt/a (following Neumann et.\ al.\cite{neumann-zeyen-ea-2023a}) --- this slashes natural gas use to 7\% of 2021 levels \cite{eurostat-2023,departmentforenergysecurity&netzerouk-2023} and renders fossil oil obsolete in the model (Supplementary Note S2).
Still, Europe overall can fully supply all energy demand from the residential, services, transportation and industry sectors using only local renewable generation, existing nuclear power and marginal use of abated natural gas --- remaining demand for natural gas can likely be supplied even by Norwegian gas fields alone \cite{oljedirektoratet-2022}.
For comparison, in 2022 the fraction of imported energy in the EU27 countries was 63\% \cite{eurostat-2024} and the goal for 2040 lies at 26--34\% \cite{europeancommission-2024}.

\begin{figure*}
    \centering
    \includegraphics{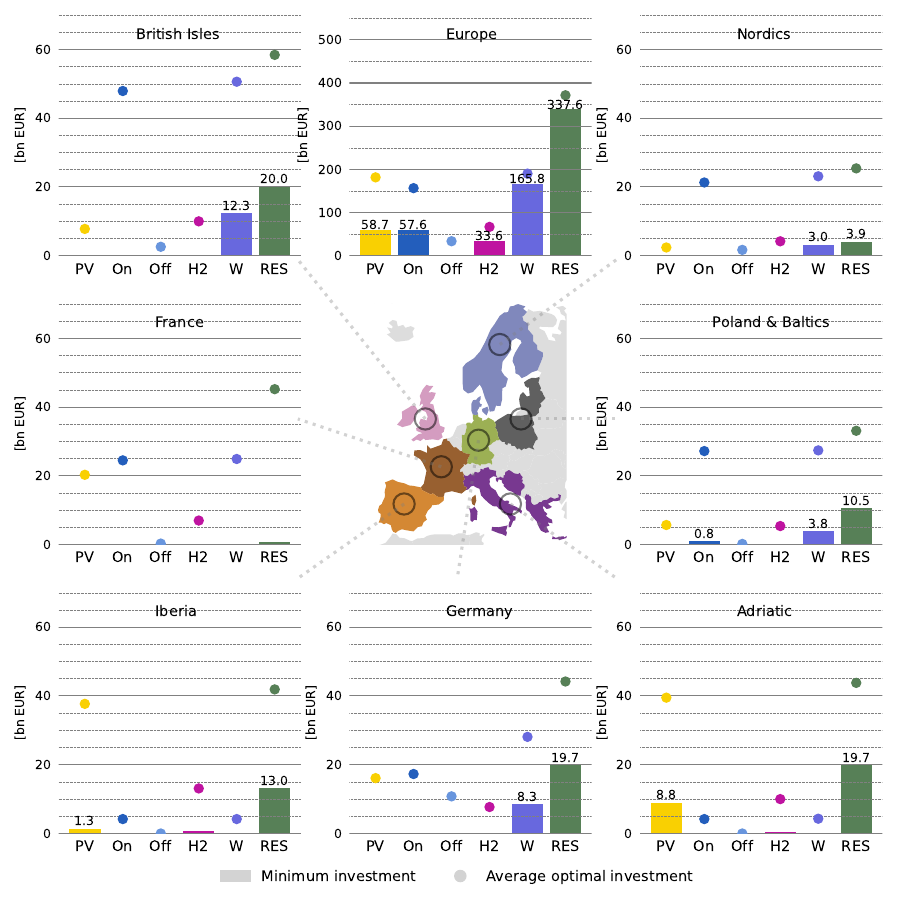}
    \caption{Comparison of minimal regional and European robust investment (in bn EUR, annualised) and average optimal investment across all scenarios. We present the investment levels in the four key technologies required for all robust designs (``PV'': solar, ``On'': onshore wind, ``Off'': offshore wind, ``H2'': hydrogen infrastructure) as well as minimal total investment in wind power (``W'': on- and offshore wind) and in renewable generation (``RES'': solar, onshore wind and offshore wind). The European investment values are the minimum across all focus regions' robust solutions (for the average optimal investments, they are the average across all scenarios and focus regions). Only indispensable investments above 1 bn EUR are annotated. For example, across robust system designs Germany requires investment of at least 19.7 bn EUR in total renewable generation, 8.3 bn EUR of which in wind power. Neither of solar, onshore or offshore wind power on its own is strictly necessary in Germany. Across the 12 scenarios (Methods), German optimal investment in renewable generation is on average 44.2 bn EUR. See Supplementary Figures S8--10 for  versions of this figure excluding the restrictive solar land-use scenarios as well as including restrictive wind land-use scenarios; minimum investment levels for wind power of any kind drop significantly for all regions as well as Europe overall if higher solar investment is allowed. For instance, Figure S8 shows that minimum overall European wind power investment drops from 165.8 to only 26.1 bn EUR when the restrictive solar land use scenarios are dropped, increasing the land available for utility solar by threefold.}
    \label{fig:investment-ranges}
\end{figure*}

At a regional level, we observe a remarkable variety of robust and cost-effective combinations of regional investments in solar power and onshore wind  (\cref{fig:overview-onwind-solar} (d)--(j)).
As a matter of fact, most focus regions can get away entirely without any onshore wind or without any solar power.
\cref{fig:investment-ranges} shows that a number of regions need a certain minimum investment in wind power of any kind, but as wind resources tend to be more abundant in coastal proximity, the balance between onshore and offshore wind can be adjusted relatively freely.
Locally reducing overall investment in renewables is also possible, though for the purposes of this study we enforce a net self-sufficiency level of 75\% for every country in the model (Methods; see Supplementary Note S3 and Figures S6, S7).
The minimum investment levels for any kind of renewable seen in \cref{fig:investment-ranges} largely reflect this constraint.
Partial self-sufficiency does not drastically increase total system cost but prevents the unnecessary exploration of system designs in which individual regions heavily rely on imports --- we consider such designs unlikely to be realised.

Although hydrogen infrastructure is crucial for decarbonisation when carbon sequestration potential is limited and needs investment within Europe of at least 34 bn EUR in all robust designs, any \emph{individual} region can forego hydrogen infrastructure entirely.
Thanks to transportation at lower cost than electricity, we find that every studied region can become a hydrogen powerhouse, accumulating tens of billions of hydrogen infrastructure investment and reducing (but not eliminating) the need for hydrogen infrastructure in the rest of the system.
Across robust system designs, we find total annual green hydrogen production in the range of about 2800--3900 TWh (comparable to an EU+UK natural gas consumption of around 4500 TWh in 2022\cite{eurostat-2023,departmentforenergysecurity&netzerouk-2023}).
The variations in total green hydrogen production reflect its potential (but not indispensable) role as a provider of operational flexibility; system cost slack levels above 5\% further increase the range.
Green hydrogen is found to be indispensable for the production of synthetic oil --- demand which determines the lower feasible bound on green hydrogen production.
Beyond that, any additional hydrogen is channelled in large part towards fuel cells, with smaller quantities also used to produce additional synthetic oil and gas.

In a similar fashion, we find robust designs in which some regions export energy in significant quantities even as all other countries cover three quarters of their annual net energy demand locally.
A striking example is provided by the Nordics, which have the potential for annual energy exports of up to 2000~TWh based on annualised investment of over 230 bn EUR ($\sim$1800~GW of onshore wind compared to currently installed 30~GW).
This is due to plentiful wind resources as well as large availability of land area for wind power --- even if the political feasibility is doubtful.
For comparison, Poland and the Baltic countries, France, and the British Isles are each found to have the potential for 1200--1300~TWh of annual exports.
A complete overview over export ranges is given in Supplementary Figure S11 (see also Supplementary Note S4); the exact values can depend on total system cost slack as well as the self-sufficiency constraint.

In contrast, Germany remains dependent on energy imports from the European grid (in line with national projections for 2050 \cite{germangovernment-2023}) in any robust design.
The dependence on imports is due on one hand to a high energy demand, and on the other hand due to limited land area available for renewable energy.
Especially our scenarios assuming reduced land area availability for solar (reduced allowable average density down to 1.7~MW/km$^2$ from 5.1~MW/km$^2$; Methods) prevent Germany from reaching full energy self-sufficiency; see Supplementary Figure S3 for an overview over robust trade-offs between regional solar and wind investment \emph{without} considering the restrictive solar land use scenarios.
This indicates that even Germany could become a net energy exporter, but only at the cost of greater potential land use conflicts.

\subsection*{Locally indispensable technologies and combinations}
Our 75\% net energy self-sufficiency constraint fixes certain minimum investment levels in renewables for some regions; others have sufficient existing low-carbon power generation.
The competitive edge due to high capacity factors make solar energy in Southern Europe and wind power particularly in the British Isles and Germany indispensable (\cref{fig:investment-ranges}).
In the case of the UK and Germany, some (onshore) wind investment is seen to be the only alternative in order to supply their significant energy demands, lacking other cost-effective low-carbon alternatives.
The minimum levels of wind power investment are significantly reduced, however, under less restrictive land-use limitations for solar (Supplementary Figure S8).  

\begin{figure*}
    \centering
    \includegraphics{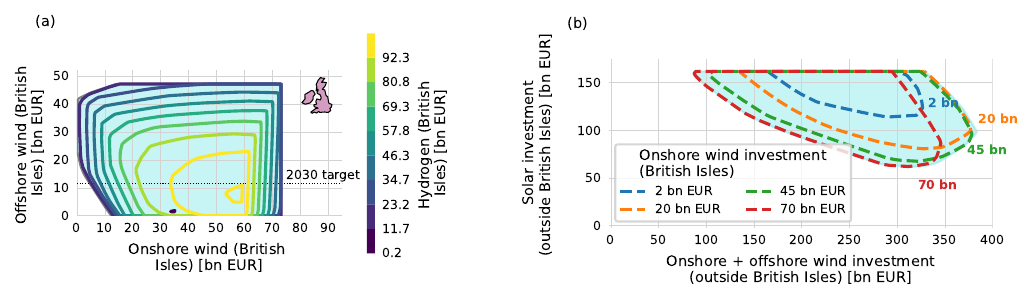}
    \caption{
    Internal dynamics between wind power and hydrogen on the British Isles (a) and the effects of British wind power on continental renewable investment (b).
    % Caption (a):
    The left panel shows that low investment in wind power on the British Isles (of at least 12.3 bn EUR as by \cref{fig:investment-ranges}) is connected with low investment in hydrogen infrastructure, as sufficient affordable electricity for green hydrogen is lacking.
    The annotated 2030 target represents the sum of UK \cite{britishgovernment-2022} and Irish \cite{irishgovernment-2023} offshore wind goals amounting to 50~GW and 5~GW respectively.
    Existing onshore and offshore capacities (at the end of 2023) are retrieved from IRENA\cite{irena-2024} and sum to 20.3~GW of onshore wind and 14.8~GW of offshore wind.
    In both cases, capacities in GW are converted to annualised investment in bn EUR using the capital cost assumptions also used elsewhere in the model.
    % Caption (b):
    The right panel shows the significant impact that British onshore wind has on robust invest levels in renewables in continental Europe.
    Renewable investment outside the British Isles is decomposed into onshore + offshore wind ($x$-axis) and solar ($y$-axis); the full space of robust solutions is shaded in turquoise.
    The dashed lines show how this design space changes depending on onshore wind investment on the British Isles.
    Very low onshore wind investment in the British Isles reduces the overall design flexibility of the rest of the system: the 2 bn EUR level shows a much smaller space.
    At the highest levels, on the other hand, the maximum viable wind investment in continental Europe is reduced.
    Supplementary Figures S12--14 are versions of this figure without the restrictive solar land-use scenarios, as well as including scenarios with limited onshore wind power land-use.
    }
    \label{fig:combined-UK-regional-vs-continental}
\end{figure*}

As a specific example, \Cref{fig:combined-UK-regional-vs-continental} (a) shows ranges of robust local investments in wind power in the British Isles.
We see that neither offshore wind nor onshore power are strictly speaking necessary; policymakers can substitute cheaper onshore wind power with less visible capacities offshore.
Low investment in hydrogen infrastructure allows vast ranges of investments in the different wind technologies (also depending on solar investments, not shown).
A strong commitment to hydrogen in the UK, however, relies on higher investment in wind power (especially onshore) and reduces the design space.
This shows that the UK could host a large hydrogen industry, which is only competitive in the presence of abundant renewable energy and relies on energy exports \cite{britishgovernment-2022}.
Overall, robust designs need at least 20 bn EUR of annualised renewable investment in the UK and Ireland, and at least 12.3 bn EUR in wind power --- a tripling of today's capacity to about 100~GW \cite{windeurope-2023} (\cref{fig:investment-ranges}).

The Nordics or France on the other hand, with significant existing capacities of hydropower and/or nuclear power, can meet 75\% of their domestic demand with fewer additional renewables, and without any single essential technology.
This is in part due to electrification (including a switch to heat pumps) and the resulting reduction in primary energy demand.
Furthermore, our implementation of the self-sufficiency constraint in terms of yearly net balance, counting energy content in terms of lower heating value for non-electricity carriers, allows e.g.\ France to export electricity and import high grade synthetic fuels, essentially outsourcing efficiency losses in hydrogen and derivative fuel production.
The observed minimal investment levels for France and Nordics (\cref{fig:investment-ranges}), though robust and cost-effective from a system perspective, are below current national targets for renewable expansion and ambitions for energy exports \cite{frenchgovernment-2023, norway-hydrogen-strategy, northsea-declaration,swedishgovernment-2023,danishgovernment-2023,finnishgovernment-2023}.

\subsection*{Effects of national policies on the European energy system}

While a patchwork of national strategies forms Europe's net-zero transition, we find that investment in some regions and technologies has a much larger impact on the success of the overall transition than others.
On one hand, lack of renewable generation in one region can lead to an inefficient overall system design, increasing the chance of continental cost overruns.
On the other hand, high regional investment can out-compete similar investments elsewhere.

For instance, the potential for onshore wind development on the British Isles is large enough to shift minimum and maximum robust renewable investment in the remainder of Europe.
\cref{fig:combined-UK-regional-vs-continental} (b) shows that for robust, cost-effective system design, lack of onshore wind power on the British Isles forces continental Europe to compensate with additional renewables (either solar or wind).

\begin{figure*}
    \centering
    \includegraphics[width=18cm]{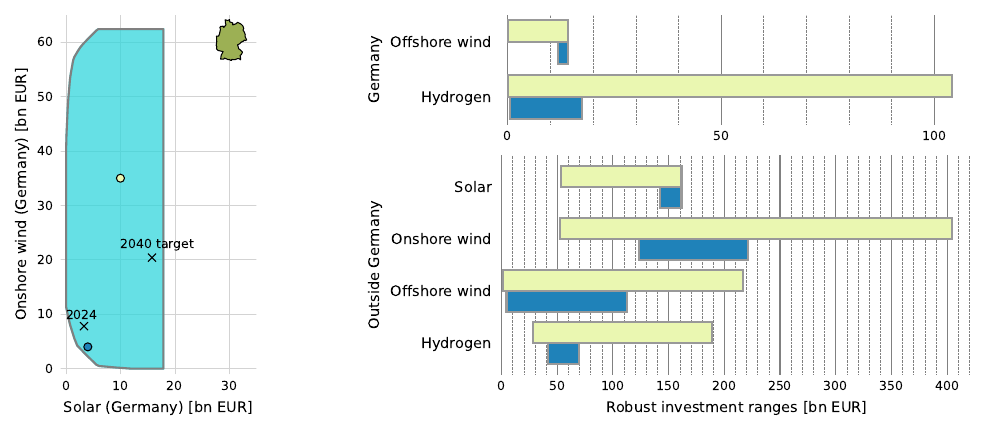}
    \caption{Left, the robust design space for solar and onshore wind inside Germany, with two particular designs / points marked as blue and yellow dots.
    Current (April 2024) \cite{arbeitsgruppeerneuerbareenergien-statistik-2024} and targeted capacities for 2040 \cite{germangovernment-2023} are marked with crosses; these are converted from GW to bn EUR annualised investment using the technology-specific capital costs also used elsewhere in the model.
    Right, robust investment ranges for remaining technologies inside Germany and all technologies under consideration outside Germany, at the investment levels in onshore wind and solar inside Germany given by the two points on the left.
    Blue bars correspond to the blue point and likewise for yellow.
    The ranges are expected to be of similar relative proportions but wider or narrower with higher or lower slack levels, respectively.
    Observe that the investment ranges for the blue point, being located closer to the boundary of the robust design space, are significantly smaller (as in \cref{fig:regional-impact-europe}).
    German as well as European policymakers have an interest in ensuring large near-optimal feasible investment ranges (i.e. a large near-optimal space), as this translates to greater design robustness.
    Supplementary Figures S15--17 are versions of this figure without the restrictive solar land-use scenarios, as well as including scenarios with limited onshore wind power land-use; both Germany and the rest of Europe have wider ranges of options when more solar can be built.
    }
    \label{fig:DE_triple-plot}
\end{figure*}

Investment decisions in some regions can also significantly restrict or enable design flexibility in the rest of the continent, leaving either many different or rather few robust options for a European net-zero energy system by 2050.
For example, \cref{fig:DE_triple-plot} shows how low renewable generation in Germany leaves little room to manoeuvre among remaining cost-effective robust solutions.
Allocating an additional 35 bn EUR/a (4\% of total system costs) to onshore wind and solar in Germany, however, frees up wide ranges of robust investment choices in the rest of Europe --- despite decreasing available capital in the rest of the system.
Mapping our near-optimal spaces thus provides policymakers with useful information on the robustness of energy system designs.

\begin{figure*}
    \centering
    \includegraphics[width=18cm]{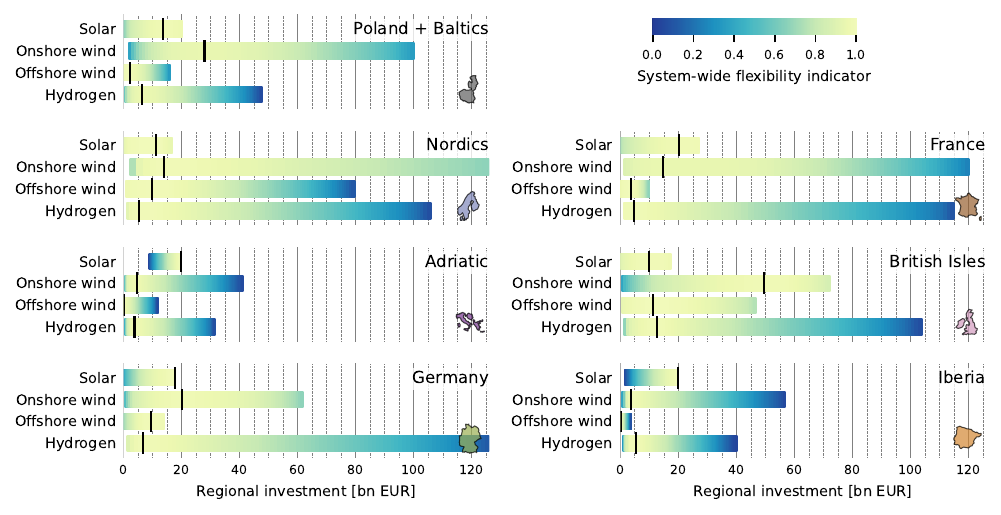}
    \caption{Effects of regional investment decisions on the design flexibility of the rest of the European system.
    The calculation of the flexibility indicator (where 1 is maximally flexible) that is plotted is explained in the Methods.
    The levels of regional investment leading to maximal outside flexibility are marked with black vertical bars.
    The outer limits of each bar represent the minimum and maximum robust regional investment levels for each technology.
    For instance, investment in solar in Poland and the Baltic countries has a positive effect on design flexibility in the rest of the system from 0 and up to about 14 bn EUR, with only a marginal decrease in flexibility upon further investment.
    The blue and yellow dots in \cref{fig:DE_triple-plot} are also examples of designs with low and high flexibility indicators, respectively.
    Robust onshore wind investments in the Nordics are cut off at 130 bn EUR (\cref{fig:overview-onwind-solar}(e)).
    While robust investment ranges widen with increasing slack levels, the investment levels maximising system-wide flexibility are not expected to change significantly under alternative slack levels.
    Supplementary Figure S18 shows the ``converse'' to this figure, with inside and outside dimensions swapped.
    Supplementary Figures S19 and S20 are more detailed versions of this figure and its converse, showing the derivative of the mean design width (proportional to the derivative of the flexibility indicator).
    See also Supplementary Note S5 for additional details on methodology.
    }
    \label{fig:regional-impact-europe}
\end{figure*}

In order to systematically quantify the effect of regional choices on the rest of the system, we introduce a system-wide design flexibility indicator based on the size of the space of robust European energy solutions subject to a given investment level in a particular region and technology (Methods).
A low score on this indicator means that the given regional investment level leaves few robust solutions for the rest of Europe, increasing the chance of energy shortage or cost overruns. 
\cref{fig:regional-impact-europe} shows the system-wide flexibility indicator across the robust investment ranges for different technologies in our seven focus regions.

Solar in Iberia and in the Adriatic region stands out as a critical piece in the European energy transition.
Lacking solar power expansion in either of these regions drastically reduces planning flexibility for the whole rest of the continent.
Other critical pieces include solar power in Germany and, to a lesser extent, onshore wind power in Poland and the Baltic countries, Germany, and the British Isles.
The disproportional importance of solar in southern Europe can be explained by its low cost (meaning that alternatives are relatively expensive and not near-optimal) and limited land availability (meaning that lacking solar power in one region cannot easily be compensated for in another region).

Other regional renewable expansion has a remarkably low impact on the planning flexibility of the rest of the system.
This is the case for solar in northern Europe as well as France, as well as onshore wind in general.

\subsection*{Effects of European policy on regional design flexibility}

Given the inter-connected nature of the energy system, regional investment decisions are sometimes also strongly impacted by policies in the rest of Europe.
We generally see two distinct effects: in instances where the rest of Europe invests heavily in (certain kinds of) renewables, most --- but not all --- individual regions' potential for becoming major energy exporters is shrunk.
If European investment is weak, some regions must compensate to achieve a robust and cost-effective system design.

\begin{figure*}
    \centering
    \includegraphics[width=18cm]{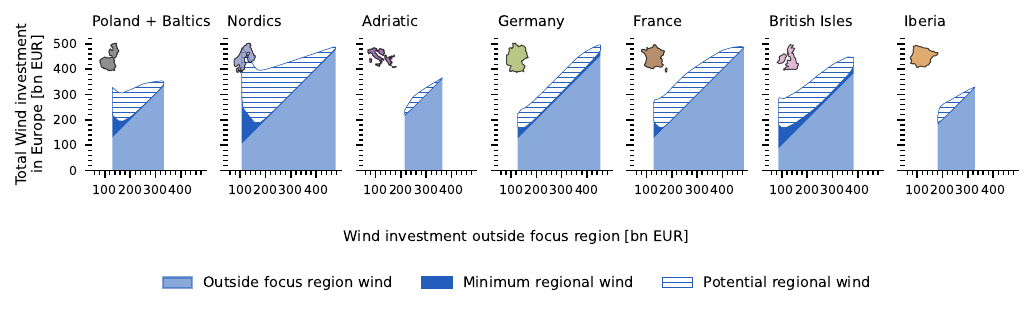}
    \caption{Minimum and maximum potential ranges of total wind power investment in different regions, as a function of wind power investment outside the respective regions. In all cases, ``wind power'' means onshore and offshore wind combined.
    Note that each subplot represents separate results from differently focused models, hence the disagreement on maximum overall viable wind power investment.
    Supplementary Figures S21 and S22 show the corresponding regional potential of solar and hydrogen investment, respectively.
    Supplementary Figures S23--28 show versions of this figure and S21 (solar) with different combinations of restrictive land use scenarios for solar and wind included; minimum regional wind investment is significantly reduced when greater land use for solar is allowed.}
    \label{fig:wind-ranges}
\end{figure*}

\Cref{fig:wind-ranges} illustrates both effects.
The massive potential for wind power development in both Poland and the Baltic countries as well as in the Nordics decreases steadily with rising wind investment in the rest of Europe.
Simultaneously, each of these two regions is individually forced to compensate strongly if the respective rest of the system falters in wind power.
This dynamic results from low domestic demand but plentiful wind resources and makes large-scale wind development generally export-dependent.
As outside developments determine what becomes over- or underinvestment, coordination of wind power expansion around the North and Baltic seas may be required in order to ensure adequate and cost-effective system designs.

Elsewhere, regions with high domestic demand are less affected by wind investment in the rest of the continent, which makes wind power less susceptible to cost-inefficient lock-ins and overinvestment:
the \emph{potential} wind investments in France, Germany and the British Isles remain stable even if wind power is expanded continent-wide (\cref{fig:wind-ranges}).
Moreover, the British Isles and especially Germany \emph{need} wind investment to cover large shares of domestic demand almost regardless of the wind development in the rest of Europe.
These regions stand apart as requiring the highest minimum investment in wind power across any robust system design (\cref{fig:investment-ranges}).
The British Isles also have the potential to export wind power and this exposes local wind investment to some competition, but these exports are never completely out-competed by continental wind investment.
Joint mapping of regional and continental robust design spaces thus provides local policymakers with a unique perspective on how exposed their energy goals are to outside developments.

\section*{Discussion}
\subsection*{Comparison to existing literature and policy}
Trade-offs and broad ranges of options have been previously shown for the European electricity \cite{neumann-brown-2021,neumann-brown-2023} sector; similar research on the \emph{energy} system (including the heating, transportation and industry sectors) is sparser \cite{pickering-lombardi-ea-2022}.
Here, we contribute with a more systematic exploration of options than previously available, showing not only extensive trade-offs on the continental level but especially on the regional level, where policymakers can plan to build significantly less of or entirely exclude one renewable energy source in favour of another.

However, the design space of individual regions does not exist in a vacuum but is strongly coupled with the surrounding system.
The integration is expected to deepen due to benefits of transmission \cite{schlachtberger-brown-ea-2017,PyPSAEurSec,horsch-brown-2017,trondle-lilliestam-ea-2020a} and hydrogen networks \cite{neumann-zeyen-ea-2023a}.
We quantify the interaction between national and continental policies, laying out how both sides can increase or decrease the extent of the design space of the other, respectively.
Under conservative land-use restrictions, each region with good solar capacity factors should contribute in order for Europe overall to reach an adequate amount of installed solar capacity; wind power and hydrogen infrastructure is less geographically constrained.
Coordinating regional long-term plans can lead to larger design spaces --- the benefits being higher robustness (lower vulnerability to uncertainty) or more flexibility and adaptability.
Particularly onshore \emph{and} offshore wind power around the North- and Baltic seas may need coordination in order to prevent over- or under-investment: the Ostend Declaration \cite{northsea-declaration} covering North Sea offshore wind and targeting at least 300~GW by 2050 is a step in the right direction.

This study reveals land use restrictions for solar power as a critical factor which cuts down the number of robust solutions (when including scenarios allowing only an average density of 1.7~MW/km$^2$ instead of 5.1~MW/km$^2$ within available land); previously, weather years have already been identified as having a major impact on the energy system design space \cite{grochowicz-vangreevenbroek-ea-2023}.
The restrictive case (with 1.2\% of the total land area of the modelling region available for utility solar) requires a minimum of about 165 bn EUR/a European wind investment ($\sim$1300~GW of onshore wind); tripling the total area available for utility solar reduces minimum wind investment to only 26 bn EUR ($\sim$200~GW of onshore wind), compensated by solar.
For comparison, the European Environmental Bureau estimates 5.2\% of the total land area of the EU to ``suitable'' for solar and wind development \cite{mayaperera-cosimotansini-2024}.
Similar land-use restrictions for wind power have a much smaller effect (Supplementary Figure S9).

Moreover, overall design flexibility is found to be uniquely sensitive to solar investment in southern Europe and Germany, with for instance the greatest benefits unlocked by around 400~GW of solar on the Iberian peninsula --- a quadrupling of the 2030 target of around 100~GW \cite{spanishgovernment-2023,portuguesegovernment-2023}.
Germany is on a comparatively good track, already targeting 400~GW of solar by 2040 \cite{germangovernment-2023}.
By allotting enough land area for utility solar power in southern Europe, policymakers can lower the risk of energy supply shortage or cost overruns.
Assessing the relative impacts on total system robustness of renewable investment in different regions is made possible by the regionally resolved near-optimal methodology we introduce.

Energy investment in individual countries and regions, while flexible, is affected by the direction of the rest of the system.
Any of the studied regions can be energy net importers (or exporters, except for Germany under restrictive solar land-use restrictions), but not all at the same time.
Our results go beyond the 10~Mt green hydrogen target for 2030 \cite{eu-hydrogen-strategy-2020,uk-hydrogen-strategy,german-hydrogen-strategy-2023,norway-hydrogen-strategy} and point at the need of about 100~Mt of green hydrogen by 2050, ($\sim$34 bn EUR/a in the overall hydrogen sector excluding electricity costs), representing around 5\% of total energy system costs by 2050 and more than the annualised capital cost of all onshore wind installations in the EU as of 2023 \cite{windeurope-2023}. 
Our results complement previous research on European green hydrogen \cite{seck-hache-ea-2022,kountouris-bramstoft-ea-2024}, by revealing the wide latitude as to \emph{where} its future infrastructure could be located. 
All in all, neither individual countries nor the EU can plan the transition to net-zero emissions without taking the tapestry of energy strategies across the European continent into account.

\subsection*{Possible limitations and future research directions}

The computational effort of approximating design spaces (450 optimisations per region and scenario) restricts our analysis to 8 key variables at a time.
Moreover, robust design spaces depend on the cost slack ($\varepsilon$) by definition and upper limits are driven by the cost slack.
Our sensitivity analysis ($\varepsilon = 2\%, 10\%, 20\%$) shows changes in size but not shape of robust design spaces (Supplementary Table S1 and Figure S1), hence revealing that trade-offs in terms of regional system design influencing the rest of Europe and vice versa (i.e. \cref{fig:combined-UK-regional-vs-continental} (b), \cref{fig:DE_triple-plot,fig:regional-impact-europe,fig:wind-ranges}) hold across different slack levels.
Minimum investments, meanwhile, mainly depend on our assumptions on partial self-sufficiency as well as renewable land-use restrictions.
Our chosen 5\% cost slack (at the conservative end compared to existing literature on near-optimal solutions\cite{neumann-brown-2021,neumann-brown-2023,lombardi-pickering-ea-2020,pickering-lombardi-ea-2022}) does not necessarily reflect willingness-to-pay of stakeholders (where stated preferences can differ from realised ones \cite{trutnevyte-2016}).
The extent of the robust design space studied here depends on the set of scenarios under which each design must be feasible, and our scenario selection is more of a starting point than a canonical selection.

Additional sensitivity analyses could include alternative emissions targets, demand response, significant changes in the European fleet of nuclear power plants, renewable energy imports, technological learning, regional differences in (static) technology costs and carbon sequestration potential among others.
Other measures for operational robustness such as rolling-horizon optimisations have not been assessed.

We focus on flexibility in reaching net-zero by 2050 based on European variable renewable energy due to sustainability concerns of biomass (here close to today's level) \cite{smith-davis-ea-2016,heck-gerten-ea-2018} and the difficulties in scaling up \ch{CO2} sequestration \cite{odenweller-ueckerdt-ea-2022,kazlou-cherp-ea-2024}.
Moreover, our study does not consider the possibility of importing ``green'' fuels such as green hydrogen and derivatives from outside Europe; the availability of such imports are likely to stay limited in the foreseeable future \cite{muller-riemer-ea-2024}.
A larger availability of either biomass, abated fossil fuels, green fuel imports or affordable nuclear power could open up additional pathways to net-zero emissions.
Especially the use of hydrogen as feedstock for synthetic fuels as well as a provider of operational flexibility might be reduced.

Increasing the spatial resolution in each of the 7 focus regions separately means that there may be slight variations in European-wide modelling results between differently focused model instances.
This could be alleviated in future research by using a uniformly high spatial resolution for the entire modelling region.
While our analysis is restricted to overnight investments, multi-horizon modelling could be of great value to policymakers.
Lastly, inviting policymakers and stakeholders to interact with near-optimal modelling tools would be a major development in participatory modelling and facilitate larger social acceptance of the energy transition.

\subsection*{Conclusions}
Each region in Europe is faced with difficult dilemmas concerning a successful transition to net-zero emissions by 2050.
We introduce for the first time a systematic exploration of the options available to individual regions, revealing a larger variety of system configurations than previously shown.
Moreover, whereas earlier studies have relied on cost optimisations or single scenarios combined with near-optimal modelling, our results hold even when accounting for different cost- and land-use scenarios.
The ranges of options presented here remain not only feasible but also within 5\% from near-optimal with respect to any of the considered scenarios, including unforeseen cost increases in wind- and solar power as a well as land-use restrictions and different weather years. 

The vast planning flexibility we demonstrate underlines the agency of single regions and countries in shaping successful long-term energy policy: if onshore wind is socially unacceptable, for example, many regions can substitute it with other technologies.
On a continent with different transition speeds \cite{mataperez-scholten-ea-2019} and levels of ambition \cite{tosatto-beseler-ea-2022}, almost all countries have the possibility to benefit from building up new industries through electricity or hydrogen exports.
However, we see that certain trade-offs between regions cannot be avoided. 
To overcome opposition to climate mitigation and strengthen a fair decarbonisation, accounting for diverging interests in a panoply of transition alternatives is possible and necessary.

\section*{Methods}
\label{sec:methods}
%TC:ignore

\subsection*{Modelling framework}
For this study, we use the open-source energy system model PyPSA-Eur-Sec 0.6.0 \cite{PyPSAEurSec} (since merged into PyPSA-Eur) which represents the European energy system including the power, heating, industry, transport and agricultural sectors.
It includes a detailed representation of the existing electricity transmission and gas networks, as well as generation sites and spatially resolved demand of different sectors.
With the aim of a 100\% $CO_2$ emission reduction in the European power, heating, transportation and industrial sectors, the model finds investment and operational decisions for generation, transmission and storage in order to minimise total annualised system costs and meet projected 2050 energy demand.
Final energy demand in the electricity, heat, transportation and industry sectors is exogenously fixed, but energy flow within the model, including electricity generation and transmission, the use of fossil fuels versus the production of synthetic fuels, the choice of heating technology and more are endogenously optimised. 
For a precise description of the model formulation see Brown et.\ al.\cite{PyPSAEurSec} and the supplemental experimental procedures of Neumann et.\ al.\cite{neumann-zeyen-ea-2023a}.

Estimates of fixed and operational costs for 2050 are taken from \url{https://github.com/PyPSA/technology-data}, a repository collecting cost data and learning curves from various sources.
All capital costs are annualised with a discount rate of 7\%; this discount rate is chosen because it is the default in PyPSA-Eur, hence making the results more directly comparable across studies.
While costs in the above repository are given in 2015 EUR, we have converted all cost data to 2023 EUR for the purposes of this study, using inflation data for the Euro area up to October 2023 \cite{eurostat-2023a}, amounting to a 24.5\% increase compared to \texttt{technology-data}.

The model is run with a partial greenfield approach, where existing transmission and gas networks (2019) as well as nuclear, biomass, and hydropower generation (2022) are included at today's capacities.
The above infrastructure was included because of its relatively long lifetime (including potential lifetime extensions), making existing capacities likely to approximately persist until 2050.
For nuclear energy, we exclude recently decommissioned power plants in Germany, but do not take into account future decommissioning plans in Belgium, Spain and Switzerland.
The gas and transmission networks may additionally be reinforced beyond today's capacities in the model; for transmission this is limited to 125\% of current levels (see sensitivity analysis in Supplementary Figure S29, Table S2, Note S6).
While the limitation on transmission expansion is set at the system-wide level, individual connections may be reinforced more.
We consistently see across scenarios and regionally focussed models (see below) that cost optimisations lead to approximately a quadrupling of the transmission capacity between Ireland and Great Britain, whereas no other capacities are more than tripled, and most are less than doubled.
Nuclear, biomass and hydropower generation, on the other hand, are entirely fixed (i.e. not subject to optimisation) and are not included in total system costs.

The main technologies whose expansion is optimised from scratch include solar, onshore wind and offshore wind generation, gas turbines, combined heat and power plants with and without carbon capture and storage (CCS), boilers, battery storage, hydrogen and heat storage, various power to X and other energy conversion technologies (electrolysis, steam methane reformation, methanation of hydrogen, ammonia and methonol production from hydrogen, liquid synthetic fuel production via the Fischer-Tropsch process, biogas), direct air capture and carbon sequestration.
Moreover, a single link per model node represents the distribution grid, connecting the high-voltage transmission grid to low-voltage electricity demand; rooftop solar (as opposed to utility solar) as well as home batteries are connected to the low-voltage nodes, incurring lower energy losses and distribution grid costs. 
\ch{CO2} is modelled endogenously, meaning that balances of \ch{CO2} in the atmosphere, in storage tanks and in permanent underground storage are kept track of for each time step.
As such, any captured \ch{CO2} can either be stored temporarily, stored permanently underground (permanent underground storage is limited to 200 Mt/a following the PyPSA-Eur-Sec default) or used as a feedstock for synthetic fuel production (synthetic methane, methanol or synthetic liquid fuel).
Within the limits of carbon capture and sequestration, fossil oil and gas may bought/imported at market prices; we do not consider green fuel imports (e.g. green hydrogen or derivatives).
The model includes all major potential components of a European energy system anno 2050, including the electricity, heating, transportation and industry sectors.
See the accompanying code and data as well as PyPSA-Eur documentation and a recent study on the potential of a European hydrogen network \cite{neumann-zeyen-ea-2023a} for more details.

We choose a spatial resolution of 60 nodes for 33 European countries, however, the exact allocation of nodes varies between the differently focused models (see below).
In order to reduce the computational burden further, we use a non-uniform, sequential time step aggregation \cite{pineda-morales-2018, kotzur-markewitz-ea-2018} setting in PyPSA-Eur with 1500 time segments (which we have tested to be more accurate than a comparable 6-hourly time resolution).

For all countries represented, we impose a 75\% net self-sufficiency constraint for annual energy demand (see Supplementary Figures S6, S7 and Note S3 for more details and sensitivity analysis).
It prevents the unnecessary exploration of technically robust system designs where individual regions are heavily dependent on imports --- we consider such designs unlikely to be realised.
The implementation of such a self-sufficiency constraint is a novelty in the context of sector-coupled energy system models for Europe; previously this was merely implemented for the electricity-only version of the model (PyPSA-Eur).
The constraint is implemented bounding the ratio between total yearly local energy production and total yearly energy imports.
Following the notation used to originally introduce PyPSA \cite{brown-horsch-ea-2018}, let $g_{n,r,t}$ be the operational decision variables for generators (indexed by bus $n$, generator $r$ and time step $t$) and $f_{\ell, t}$ be the operation decision variables for branch components, both passive (i.e. AC transmission lines) and active (DC transmission as well as transfer of other quantities than electricity between buses).
Let $\mu_{\ell, t}$ be the efficiency of each branch component.
All quantities tracked and transferred in a PyPSA-Eur model are energy carriers (electricity, natural gas, hydrogen, oil, etc.) except \ch{CO2}; the unit for energy is always MWh, measured in lower heating value for thermal energy carriers.
For a country $C$, let 
\begin{equation*}
    \small
    \alpha_{C,\ell,t} = 
    \begin{cases}
        -1, & \text{if branch $\ell$ starts inside $C$ and ends outside of $C$,} \\ & \quad \text{and the bus that $\ell$ ends at tracks an energy carrier,} \\
         \mu_{\ell, t}, & \text{if it start outside of $C$ and ends inside $C$,} \\ & \quad \text{and the bus that $\ell$ ends at tracks an energy carrier,} \\
         0, & \text{otherwise.} \\
    \end{cases}
\end{equation*}
Then the net yearly imports of country $C$ are measured by
\begin{equation}
    I_C = \sum_{\ell, t} \alpha_{C, \ell, t} f_{\ell, t}.
\end{equation}
Letting
\begin{equation*}
    \delta_{C,n} = 
    \begin{cases}
        1, & \text{if bus $n$ is located inside country $C$} \\
        0, & \text{otherwise.}
    \end{cases}
\end{equation*}
be the indicator function of country $C$ for buses, the total yearly amount of energy produced inside country $C$ is
\begin{equation}
    L_C = \sum_{n, r, t} \delta_{C, n} g_{n, r, t}.
\end{equation}
Then, a self-sufficiency degree of $\gamma$ for country $C$ can be ensured by adding the following linear constraint to the capacity expansion formulation:
\begin{equation}
    I_C \leq \frac{1 - \gamma}{\gamma} L_C,
\end{equation}
provided $\gamma > 0$.
The actual implementation used in this paper is slightly complicated by the presence of multi-branch components (i.e. links connecting more than two buses, controlled by a single operational decision variable per time step), but equivalent to the above.

It should be noted that the 75\% self-sufficiency constraint has little impact on cost-optimal solutions (Supplementary Figure S6), meaning that unrestricted cost-optimal results are already close to having countries with a 75\% yearly net self-sufficiency (in the sense that total system cost is close).
As such, the constraint mainly impacts the near-optimal space, preventing the exploration of solutions where individual regions become dependent on imports to a large extent.

Apart from the self-sufficiency constraint, we refer to the publication introducing PyPSA \cite{brown-horsch-ea-2018} for an exact mathematical formulation of the full optimisation problem.

\subsection*{Regional studies}
Our results are based on studying the joint space of robust system designs of the European energy system, reduced to total investment in four key technologies both inside and outside 7 selected focus regions.
For the approximations of these 7 robust design spaces, we use the same model except that we distribute the spatial nodes of the model differently in order to allocate more spatial resolution to each respective focus region and neighbouring countries.
Our focus regions are
\begin{itemize}
    \item Poland and Baltic countries (EE, LT, LV, PL) with 20+25+15 nodes,
    \item Nordics (DK, FI, NO, SE) with 20+25+15 nodes,
    \item Adriatic (AL, BA, GR, HR, IT, MT) with 30+20+10 nodes,
    \item Germany (DE) with 20+25+15 nodes,
    \item France (FR) with 20+25+15 nodes,
    \item British Isles (IE, UK) with 15+25+20 nodes,
    \item Iberia (ES, PT) with 20+15+25 nodes,
\end{itemize}
with the corresponding number of nodes inside the region, for neighbouring countries and for remaining countries (further removed than distance 1), respectively, all adding up to 60 nodes.
See Supplementary Figure S30 for a visual representation of the networks.

Supplementary Figures S31 and S32 show the impact on the intersection of near-optimal spaces of the focussed network clustering (albeit on a model with low temporal resolution.)
Overall, the impact of focussed clustering is relatively minor.

\subsection*{Scenario selection}
The factors we vary across scenarios and which we consider most impactful to modelling results include weather years \cite{zeyringer-price-ea-2018, grochowicz-vangreevenbroek-ea-2023}, cost assumptions \cite{neumann-brown-2023}, and land use \cite{patankar-sarkela-basset-ea-2023} availability.
Following Grochowicz et al. \cite{grochowicz-vangreevenbroek-ea-2023}, we select three difficult weather years (1985, 1987, 2010) and thus are particularly well-suited for resilience considerations.
We define three cost scenarios, one ``baseline'' scenario with standard cost assumptions, and two scenarios with higher capital expenditure costs for solar PV (39\% more expensive) and wind power (24\% more expensive) respectively; these ranges are taken from cost projections by the Danish Energy Agency \cite{danishenergyagency-2016}.
Points in the robust design space (i.e.\ the intersection of the near-optimal spaces arising from the different scenarios) are defined in terms of investment levels in solar, onshore- \& offshore wind and hydrogen infrastructure.
This means that a single robust design with, say, $x$ bn EUR investment in solar, would have different total solar capacities in GW in scenarios where solar has different capital costs.

For the last three scenarios (one for each weather year), we restrict the available land area for utility solar PV in comparison to the standard assumptions in PyPSA-Eur-Sec to one third: from a maximum average density of 5.1~MW/km$^2$ (the default) to 1.7~MW/km$^2$ within available land after excluding unsuitable areas \cite{PyPSAEur}.
Assuming an installation density of 60~MW/km$^2$ for utility solar plants, this corresponds to a maximum land use of 8.5\% and 2.8\% within suitable areas, respectively.
Across the modelling region, 43.6\% of the total land area is left after initial exclusions, meaning that the 2.8--8.5\% of available land area amount to approximately 1.2--3.7\% of the total land area of the modelling region.
This can be compared to a recent report by the European Environmental Bureau \cite{mayaperera-cosimotansini-2024} estimating 5.2\% of the EU's land area to be ``suitable'' for solar and wind development.
The restricted land availability scenario inhibits expansion of solar PV significantly: for instance, Germany's installation potential is capped at approximately 440~GW while the German government is targeting installed capacities of 400~GW until 2040 \cite{germanbundestag-2022}.

All in all, this gives us $12$ scenarios, as we pair the three different weather years with the three cost scenarios and the restriction on land use for utility solar ($3 \cdot (3 + 1) = 12$).

In the Supplementary Information, we also explore results with the solar land-use scenarios removed, and with scenarios where wind land-use is restricted instead.
Initial exclusions \cite{PyPSAEur} leave 74.1\% of the total land area of the modelling region available for wind power development; by default PyPSA-Eur-Sec restricts wind power land-use to 30\% of this remaining area at a density of 10~MW/km$^2$, resulting in an overall density of 3~MW/km$^2$.
In the restrictive wind power land-use scenarios used in the Supplementary Information, this is further reduced to a third (similar to solar) for a maximum installation density of 1~MW/km$^2$ or a maximum land-use of 10\% (at an installation density of 10~MW/km$^2$) after initial exclusions.

For offshore wind, certain areas and shipping corridors are also excluded outright \cite{PyPSAEur} and a maximum overall installation density of 2~MW/km$^2$ (corresponding to a maximum of $20\%$ land-use at a 10~MW/km$^2$ offshore wind farm density) is assumed.
These restrictions are not explored any further in separate scenarios.

\subsection*{Near-optimal spaces}
The near-optimal space for each of the above scenarios and focus regions, consists of feasible, alternative solutions to a cost-optimal solution which can be preferable over the cost optimum for other reasons. 
Energy system optimisation models are usually formulated mathematically as linear programs of the form $\min cx \text{ subject to } Ax \leq b$ in which case the \emph{design space}, more formally \emph{$\varepsilon$-near-optimal feasible space} \cite{decarolis-2011,neumann-brown-2021,grochowicz-vangreevenbroek-ea-2023}, is defined as $\mathcal{F}_{\varepsilon} = \{x \in \mathbb{R}^n \mid Ax \leq b \text{ and } cx \leq (1 + \varepsilon) \cdot c^*\}$ where $c^*$ is the optimal objective (minimum cost) of the original linear program.

For the purposes of this study, we use a cost slack of $\varepsilon = 5\%$. 
See the Supplementary Note S1 for a sensitivity analysis on the cost slack.
However, for each focus region we compute a uniform cost bound for all scenarios based on the optimum system cost for the most expensive scenario (roughly following \cite{grochowicz-vangreevenbroek-ea-2023}).
More precisely, let $r$ be a region and $S$ a set of scenarios, such that we get a linear program $A_{r,s} x \leq b_{r,s}$ with objective $c^*_{r,s}$ for each $s \in S$.
Then we define the $\varepsilon$-near-optimal space for scenario $s$ as
\begin{equation*}
    \mathcal{F}_{\varepsilon}^{r,s} = \{x \in \mathbb{R}^n \mid A_{r,s}x \leq b_{r,s} \text{ and } c_{r,s}x \leq (1 + \varepsilon) \cdot \max_{s' \in S} c^*_{r,s'}\}
\end{equation*}
Defining near-optimal spaces for a set of scenarios as above allows for more direct comparisons, since all spaces are defined with respect to same absolute bound on total system costs.

In models of similar size to PyPSA-Eur-Sec (with $n$ significantly greater than $10^6$), the near-optimal spaces which can be analysed effectively are projections onto a few key dimensions from the high-dimensional space $\mathcal{F}_{\varepsilon}^{r,s}$.
This is because an accurate approximation of such a near-optimal space is computationally demanding (one vertex is obtained through one optimisation of the linear program) in high dimensions \cite{pedersen-victoria-ea-2021,grochowicz-vangreevenbroek-ea-2023}.
Following previously introduced notation \cite{grochowicz-vangreevenbroek-ea-2023}, we consider the reduced, low-dimensional near-optimal space $\mathcal{A}^{r,s}_{\varepsilon} \subset \mathbb{R}^k$; related to the full-dimensional space by a linear map $\sigma \colon \mathcal{F}_{\varepsilon}^{r,s} \to \mathcal{A}_{\varepsilon}^{r,s}$.
In our case, we map down to $k = 8$ key dimensions: the total investments \emph{inside} and \emph{outside} the focus region in utility solar, onshore wind, offshore wind, and hydrogen infrastructure. 
Hydrogen infrastructure investments consist of investment in electrolysis, fuel cells, pipelines (both new and retrofitted from existing gas pipelines), synthetic methane and fuel production from hydrogen (methanation, Fischer-Tropsch process respectively) and steam methane reforming plants (with or without carbon capture) as well as hydrogen storage.

For a more detailed description of the approximation methodology and validation see \cite{grochowicz-vangreevenbroek-ea-2023}; in this case, we conduct 450 optimisations for a satisfying approximation of each near-optimal space.
See Supplementary Figures S33 and S34 for a visualisation of the individual near-optimal spaces as well as their intersection for one of the focus regions.

It can be instructive to contrast the above approach with a global sensitivity analysis.
In a global sensitivity analysis, points are sampled from a parameter space (with individual parameters representing costs or other model input parameters), and a cost-optimal solution to the energy system model is computed for each sampled point in the parameter space.
The result is likewise a collection of different system designs, each feasible and optimal with respect to its corresponding parameters.
However, the set of solutions obtained through a global sensitivity analysis is not necessarily representative of the entire space of near-optimal solutions of a baseline model; they are not guaranteed to be near-optimal either (with respect to some baseline model).
Thus, while a global sensitivity analysis is useful in order to study the effects of various parameters on cost-optimal system design, mapping out near-optimal spaces is a more suitable method when the goal is to characterise the variety of solutions that are available at a moderate cost increase.

\subsection*{Robust solutions}
To find cost-effective solutions that are feasible (and near-optimal) notwithstanding the uncertainties encoded in our selection of scenarios, we look for investment decisions that lie in the near-optimal space for each scenario.
We formally define the space of \emph{robust solutions} (or robust design space) as points in $\mathbb{R}^k$ that lay within the intersection of the reduced near-optimal spaces for every scenario $s \in S$ under consideration:
\begin{equation*}
    \mathcal{I}_{\varepsilon}^r = \bigcap_{s \in S} \mathcal{A}_{\varepsilon}^{r,s}.
\end{equation*}
We refer to \cite{grochowicz-vangreevenbroek-ea-2023} for an extensive overview of the intersections of near-optimal feasible spaces with all due details.

\subsection*{The flexibility indicator}
For \cref{fig:regional-impact-europe} we introduce a flexibility indicator based on the \emph{mean width} of the robust design space (defined in more detail below).
The mean width of the robust design space in the 4 outside investment dimensions is a simple metric for the variety of robust solutions and serves as a quantification of design flexibility of policymakers.
The values between 0 and 1 for each regional investment dimensions show which investment levels allow for most robust solutions from a continental perspective.
For instance, investing ca. 5 bn EUR/a in onshore wind in the Adriatic gives the rest of Europe the largest flexibility, whereas onshore wind investment in the order of 40 bn EUR/a --- while still being robust --- restricts the alternatives for the remainder of the system.
In the Supplementary information, we present additional figures (S19 and S20) for the derivative of mean width, which can identify at which investment levels most flexibility can be gained or lost. 

For a precise mathematical definition, consider the space of robust designs $\mathcal{I}^r_{\varepsilon}$ for focus region $r$.
Let $A = \{a_1, a_2, a_3, a_4\}$ and $B = \{b_1, b_2, b_3, b_4\}$ be the sets of inside and outside key dimensions, respectively, i.e.\ $a_1, a_2, a_3, a_4$ are investment in solar, onshore wind, offshore wind and hydrogen infrastructure in bn EUR inside the focus region $r$, respectively (and similarly for $B$ outside the focus region).
Now, let
\begin{equation}
    \mathcal{I}^r_{\epsilon,a_i = c} = \{x \in \mathcal{I}^r_{\varepsilon} \mid a_i = c\} \quad \text{where } x = (a_1, a_2, a_3, a_4, b_1, b_2, b_3, b_4)
\end{equation}
be the subset of $\mathcal{I}^r_{\varepsilon}$ for which $a_i$ equals a constant $c$.
Then we define the \emph{mean width of $\mathcal{I}^r_{\varepsilon}$ at $a_i = c$} as
\begin{equation}
    d^r_i(c) = \frac{1}{|B|} \sum_{b \in B} \left(\max \{b \mid x \in \mathcal{I}^r_{\varepsilon, a_i = c}\} - \min \{b \mid x \in \mathcal{I}^r_{\varepsilon, a_i = c}\}\right),
\end{equation}
where we use the convention that $\min \emptyset = \max \emptyset = 0$.
That is, $d^r_i(c)$ is the mean of the widths of $\mathcal{I}^r_{\varepsilon, a_i = c}$ in each of the outside dimensions $B$, and equals $0$ when $\mathcal{I}^r_{\varepsilon, a_i = c} = \emptyset$.
Finally, we define the flexibility metric $f^r_i$ as the normalisation of $d^r_i$, namely
\begin{equation}
    f^r_i(c) = \frac{d^r_i(c)}{\max_{c \in \mathbb{R}} d^r_i(c)}.
\end{equation}
This is what is plotted in \cref{fig:regional-impact-europe} for each combination of region $r$ and inside dimension $a_i$; we cut the plots off where $f^r_i(c) = 0$.
The converse (the flexibility metric for regional investment based on fixed continental investment) is presented in Figure S6.

\section*{Data availability}

Links to data used in this study are available at \url{https://github.com/koen-vg/enabling-agency/tree/v0}; all data used are open (various licenses).

\section*{Code availability}

The code to reproduce the results of the present study is available at \url{https://github.com/koen-vg/enabling-agency/tree/v0}.
All code is open source (licensed under GPL v3.0 and MIT).

\section*{Acknowledgements}

A.G., F.E.B. and M.Z. acknowledge funding by UiO:Energy and Environment (SPATUS).

ERA5 reanalysis data \cite{hersbach-bell-ea-2018} were downloaded from the Copernicus Climate Change Service (C3S) \cite{c3s-2023}.

The results contain modified Copernicus Climate Change Service information 2020. Neither the European Commission nor ECMWF is responsible for any use that may be made of the Copernicus information or data it contains.

\section*{Author contribution statement}

Conceptualisation: all authors; Methodology: KvG and AG; Coding and analysis: KvG and AG; Writing --- Original Draft: KvG and AG; Writing --- Review \& Editing: all authors; Supervision: MZ and FEB.

\section*{Declaration of interests}

The authors declare no competing interests.

%TC:endignore

\end{document}

\listoffigures

\listoftables

\listofmynotes

\newpage

\begin{figure*}[h!]
  \centering
  \includegraphics{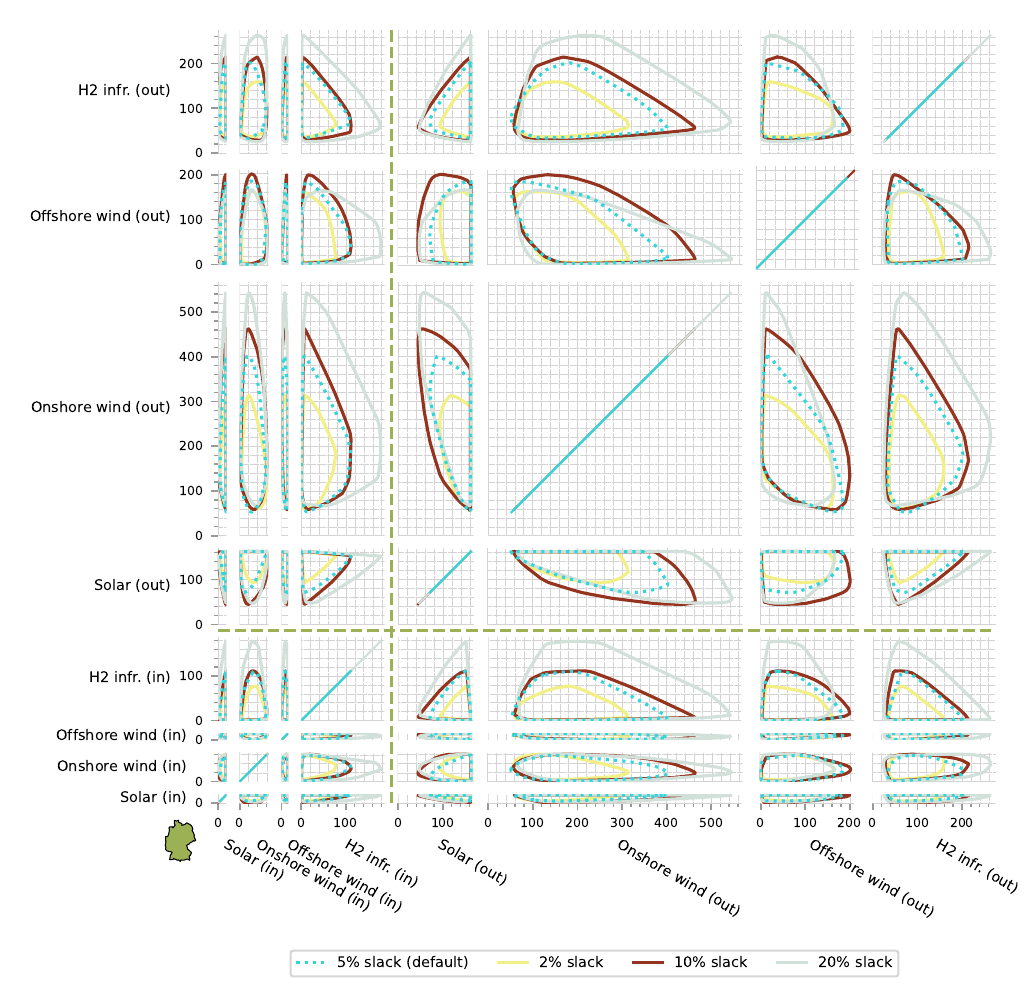}
  \caption[Intersection of near-optimal spaces for different slack levels]{
    Projections of the intersection of near-optimal spaces for the German-focused model with different slack levels (2\%, 10\%, 20\% compared to the 5\% we use in the study).
    Here, the ``(in)'' and ``(out)'' suffixes indicate that the respective dimensions denote investment inside and outside the given focus region (in this case, Germany).
    All four intersections are based on 150 iterations per near-optimal space (for the 5\%-near-optimal spaces we only took the first 150 of the 450 iterations used in the study).}
  \label{fig:slack-sensitivity}
\end{figure*}

\begin{mynote}[h!]
  \caption[Impacts of different slack levels/near-optimality constraints.]{
    \textbf{Impacts of different slack levels/near-optimality constraints.}
    In the present study, we use a near-optimality constraint (slack level) of 5\% on top of the most expensive of the 12 investigated scenarios. We have conducted a sensitivity analysis for one of the focus regions (Germany) to capture how the robust design spaces changes depending on the slack levels of 2\%, 5\%, 10\%, and 20\%.
    Different slack levels could represent an additional willingness to pay for or invest in energy-related infrastructure beyond the cost-optimal level.
    \Cref{tab:slacks} presents an overview over how key results depend on the slack level.
    \Cref{fig:slack-sensitivity} shows that although the size of the robust design spaces increases, the shape of the projections remains similar. This indicates that the trade-offs and insights presented in the main text are robust throughout different slack levels. Note that with increasing slack, the nested intersections increase in size and therefore the quality of the approximation decreases (as more than the constant 150 iterations would be needed to account for this). This explains why the approximated robust design space based on 20\% slack does not contain the other robust spaces fully --- in theory, it should.
  }  
\end{mynote}

\begin{table*}[h!]
  \centering
  \begin{tabular}{l *{4}{r@{\,--\,}l}}
    \toprule
    European robust & \multicolumn{2}{c}{2\% slack} & \multicolumn{2}{c}{5\% slack (default)} & \multicolumn{2}{c}{10\% slack} & \multicolumn{2}{c}{20\% slack} \\
    investment [bn EUR] & \multicolumn{2}{c}{($\sim~15.6$~bn~EUR)} & \multicolumn{2}{c}{($\sim~39$~bn~EUR)} & \multicolumn{2}{c}{($\sim~78$~bn~EUR)} & \multicolumn{2}{c}{($\sim~156$~bn~EUR)} \\
    \midrule
    Solar & 98.0 & 179.5 & 58.7 & 179.5 & 45.3 & 179.5 & 49.6 & 180.5 \\
    Onshore wind & 64.9 & 374.5 & 57.6 & 315.7 & 59.7 & 523.4 & 68.3 & 604.0 \\
    Offshore wind & 5.0 & 176.2 & 0.1 & 158.6 & 3.1 & 213.5 & 1.5 & 179.1 \\
    Hydrogen infrastructure & 37.5 & 236.1 & 33.6 & 118.6 & 29.8 & 326.1 & 25.5 & 441.0 \\
    Wind investment & 181.3 & 381.9 & 165.8 & 329.2 & 167.2 & 517.8 & 152.0 & 586.6 \\
    Renewable generation & 353.5 & 521.1 & 337.6 & 461.0 & 336.3 & 612.3 & 325.7 & 686.5 \\
    \bottomrule
  \end{tabular}
  \caption[Sensitivities of European robust investment to chosen slack level]{Sensitivities of European robust investment to chosen slack level. The robust investment consists of the investment inside and outside the focus region. Note that the values for 2\%, 10\%, and 20\% are based on 150 iterations with Germany as the focus region, whereas the 5\% values are consistent with Figure 2 in the main text. The values in the robust range for 5\% are minimal across all 7 focus regions and are based on 450 iterations. Note that as the slack level (and therefore the robust design space) increases, the quality of the approximation decreases. This explains the counterintuitive lower limits of solar and onshore wind and upper limit of offshore wind for 20\% slack. If approximated in sufficient detail, these spaces are nested, i.e. the robust space for 20\% contains the one for 10\%.
  }
  \label{tab:slacks}
  % \end{adjustwidth}
\end{table*}

\begin{figure*}[h]
  \centering
  \includegraphics{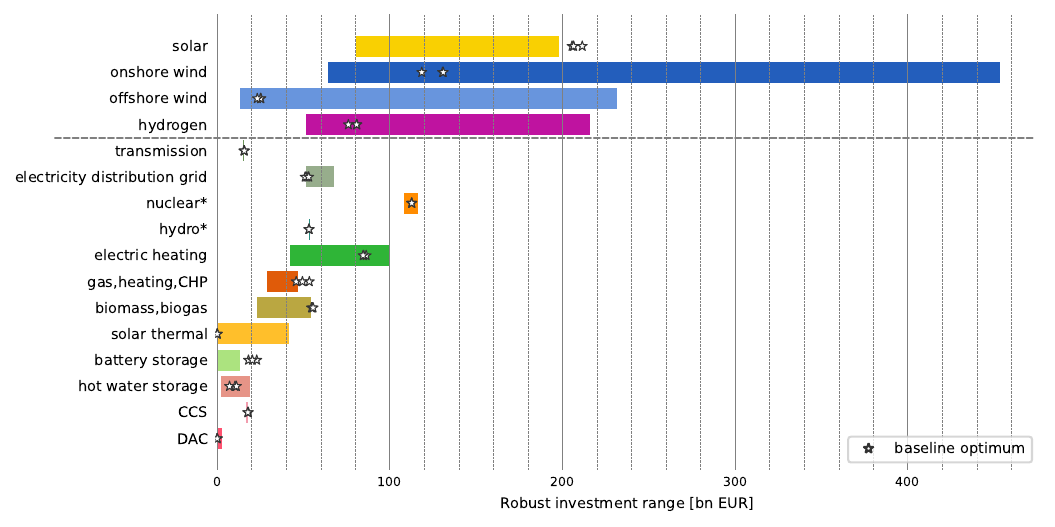}
  \caption[Total system cost distribution by category or technology]{
    Total system cost distribution by category or technology.
    The cost ranges in this plot have been calculated on the basis of optimisations of 300 robust system designs with total inside- and outside solar, onshore wind, offshore wind and hydrogen investment fixed at random vertices on the boundary of the robust design space; the bars show the minimum and maximum cost of each component or group of components shown across the 300 systems.
    Overall optimal system costs in our different focus models and scenarios (before near-optimal studies) have a range of $734$--$880$ bn EUR (not including existing transmission, nuclear and hydropower infrastructure; Methods) --- this figure shows costs in the German-focussed model.
    Comparable costs found in the baseline cost-optimal solutions (for weather years 1985, 1987, 2010) are shown with white stars.
    The top four categories represent the dimensions which have been explored explicitly --- in this figure ``inside'' and ``outside'' costs have been aggregated.
    The two starred categories, nuclear and hydro(power), have been included in this figure, but represent capacities that are not subject to optimisation and rather included at today's levels.
    There is nevertheless some variation in nuclear costs since marginal costs in the form of uranium supply can vary depending on operations, which are optimised, and included in the nuclear category.
  }
  \label{fig:total-costs}
\end{figure*}

\clearpage

\begin{mynote}
  \caption[Representation of fossil fuels and the carbon cycle]{
    \textbf{Representation of fossil fuels and the carbon cycle.}
    We strictly enforce net-zero emissions in all model runs.
    In order to achieve this, some use of negative emissions is necessary.
    However, optimal system design can depend strongly on assumed carbon sequestration potential.
    In this study, we assume a yearly sequestration capacity of 200 Mt \ch{CO2}, following \cite{neumann-zeyen-ea-2023a}.
    This limit on sequestration effectively limits the use of fossil fuels in the model.

    Natural gas use in Europe amounted to approximately 5271 TWh in 2021, with 4417 TWh in the EU27 area \cite{eurostat-2023} and 854 TWh in the UK \cite{departmentforenergysecurity&netzerouk-2023} (using gross calorific values).
    In all robust model solutions, we see a total natural gas use of around 363 TWh in the modelling region (which includes Switzerland, Norway and the Balkan in addition to the EU and the UK).
    This is about 6.9\% of the EU+GB natural gas demand in 2021.
    The use of fossil oil is not found to be competitive in any robust model solution.

    Natural gas and oil as well as non-feedstock emissions are the only way for carbon to enter the modelled cycle, and sequestration is the only way to leave the cycle. Since \ch{CO2} sequestration potential as well as non-feedstock emissions (mainly from the cement industry) are fixed in the model, this also fixes the total amount of fossil gas and oil that may enter the system.}
\end{mynote}

\clearpage

\begin{figure*}[h!]
  \centering
  \includegraphics{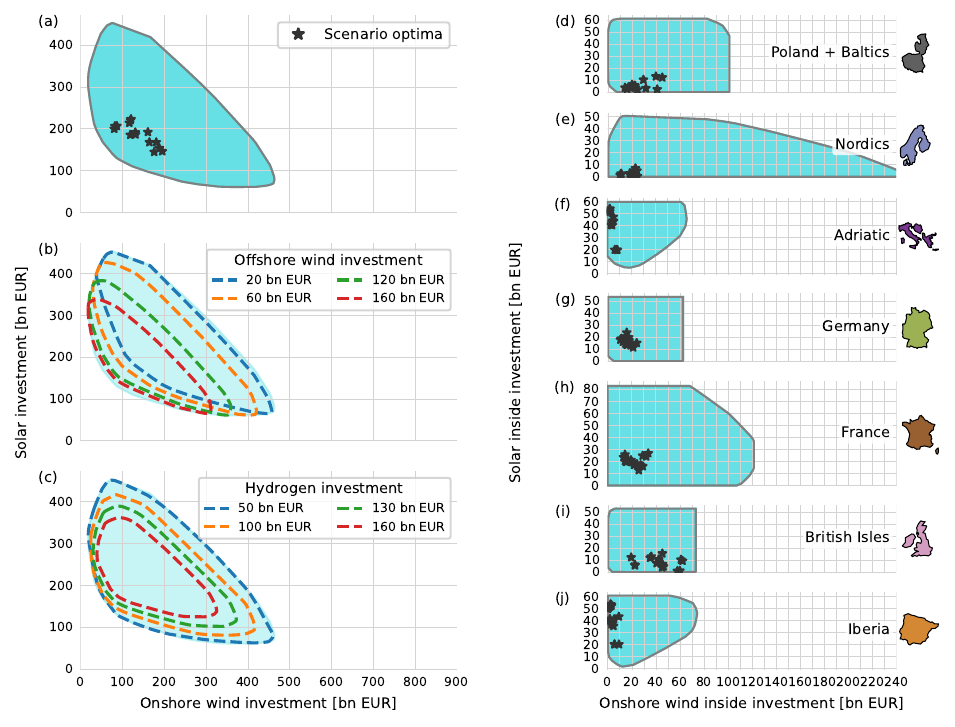}
  \caption[Version of Figure 1 without solar land-use scenarios]{Robust trade-offs between annualised onshore wind and solar investment in Europe, excluding the restrictive land-use scenarios. This parallels Figure 1 in the main text, but greater investment in solar is enabled by excluding solar land-use scenarios. As such, this figure shows projections of intersections of the near-optimal spaces corresponding to 9, not 12 different scenarios (Methods). The excluded scenarios allow only a maximum average installation density of 1.7~MW/km$^2$ within available land for utility solar; in the scenarios in this figure, the maximum such density is 5.1~MW/km$^2$.}
  \label{fig:composite--without-land-use}
\end{figure*}

\begin{figure*}[h!]
  \centering
  \includegraphics{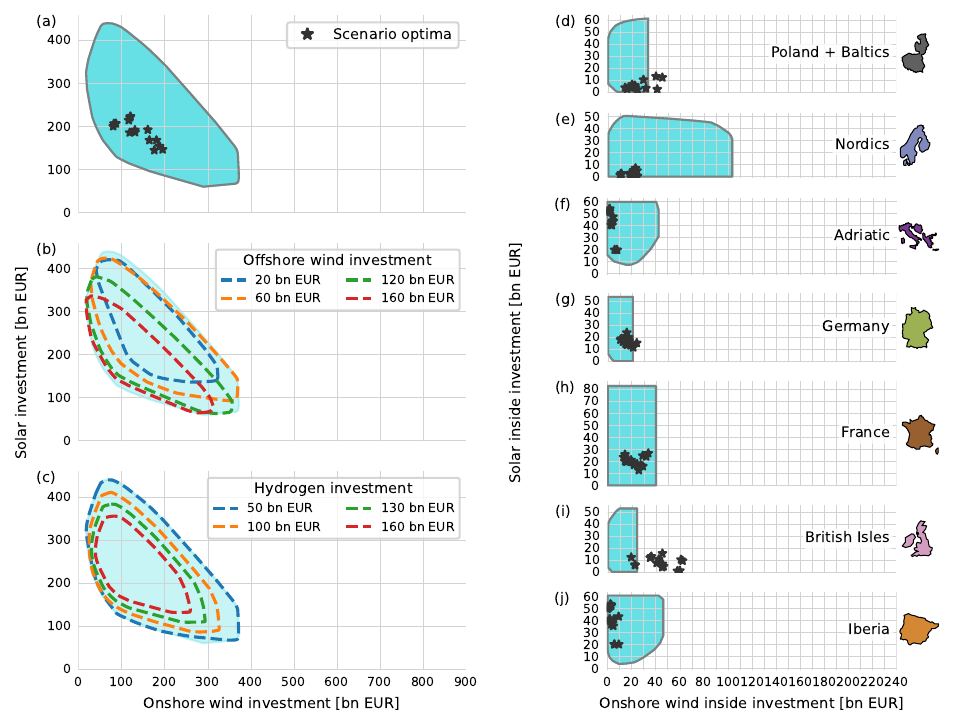}
  \caption[Version of Figure 1 with restrictive land-use for wind instead of solar]{Robust trade-offs between annualised onshore wind and solar investment in Europe, with restrictive land-use for wind instead of solar. This parallels Figure 1 in the main text, but instead of including scenarios where solar land-use is restricted to a third of the baseline, wind power land-use is restricted to a third of the baseline instead.}
  \label{fig:composite-without-solar-land-use}
\end{figure*}

\begin{figure*}[h!]
  \centering
  \includegraphics{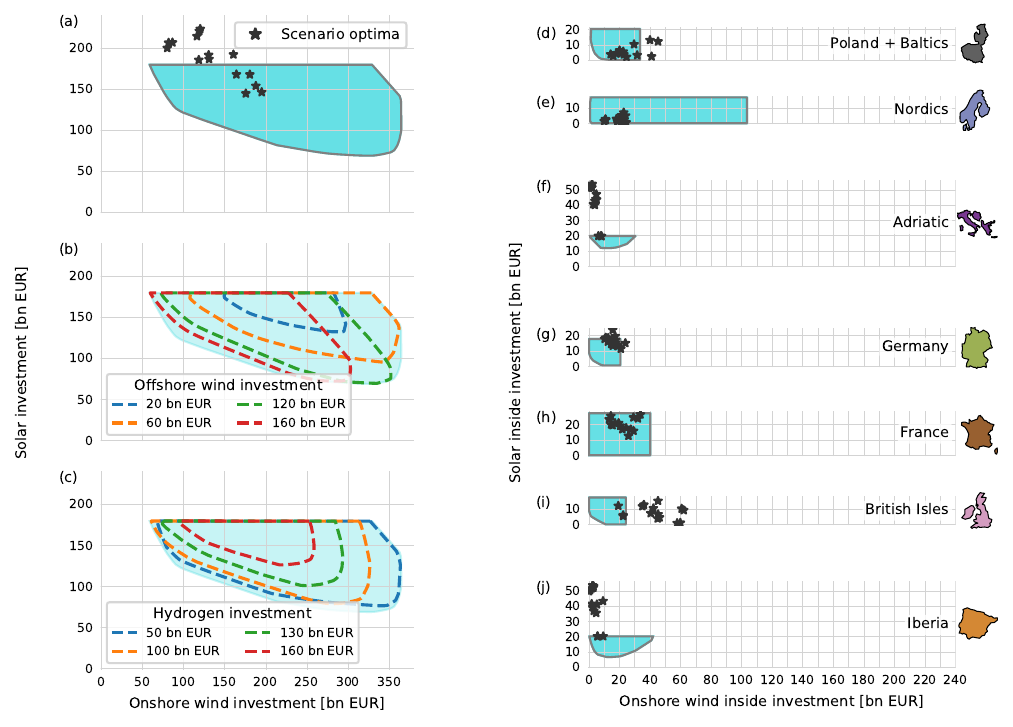}
  \caption[Version of Figure 1 with restrictive land-use for both wind and solar]{Robust trade-offs between annualised onshore wind and solar investment in Europe, excluding the restrictive land-use scenarios. This parallels Figure 1 in the main text, but in addition to the scenarios limiting solar land-use by a third, we also include scenarios that limit wind power land-use by a third.}
  \label{fig:composite-with-wind-land-use}
\end{figure*}

\clearpage

\begin{mynote}[h!]
  \caption[Impacts of self-sufficiency constraint]{
    \textbf{Impacts of self-sufficiency constraint.}
    In the present study we introduce for the first time a self-sufficiency constraint to PyPSA-Eur-Sec; this constraint having originally only been implemented for PyPSA-Eur (the electricity-only version of the model).
    The self-sufficiency is accounted for on a yearly basis, that is, a self-sufficiency level of $x\%$ means that that total yearly energy production of every country must equal at least $x\%$ of the total yearly energy demand of that country (see precise definition in the Methods).
    Below, we present a simple sensitivity analysis of total system cost with respect to given self-sufficiency levels.
    We see that requiring self-sufficiency levels above 90\% gets increasingly expensive; we choose a level of 75\% for the present study.

    Furthermore, we have conducted a sensitivity analysis for one of the focus regions (Germany) to investigate the impact of the self-sufficiency constraint (75\%) on the robust design spaces. Note that the robust design spaces in \cref{fig:self-suff-sensitivity} are smaller when dropping the self-sufficiency constraint: the total system costs are lower when the constraint is not enforced, which reduces the upper bound on near-optimality from the definition of the design spaces ($\mathcal{F}_{\varepsilon}^{r,s}$ in Methods). The reduction in available capital has a larger impact on the robust solution space than being able to rely more on imports.}
\end{mynote}

\begin{figure*}[h!]
  \centering
  \includegraphics{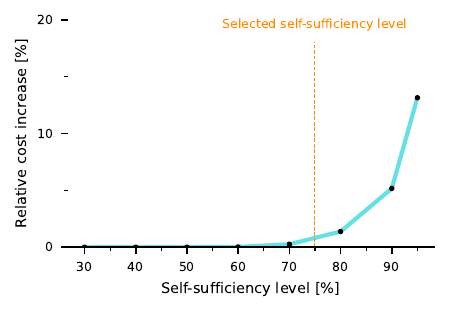}
  \caption[Optimal system cost as a function of self-sufficiency level]{Relative increase of total system costs of an optimal network depending on different levels of net self-sufficiency (see Methods).}
  \label{fig:self-suff-levels}
\end{figure*}

\begin{figure*}[h!]
  \centering
  \includegraphics{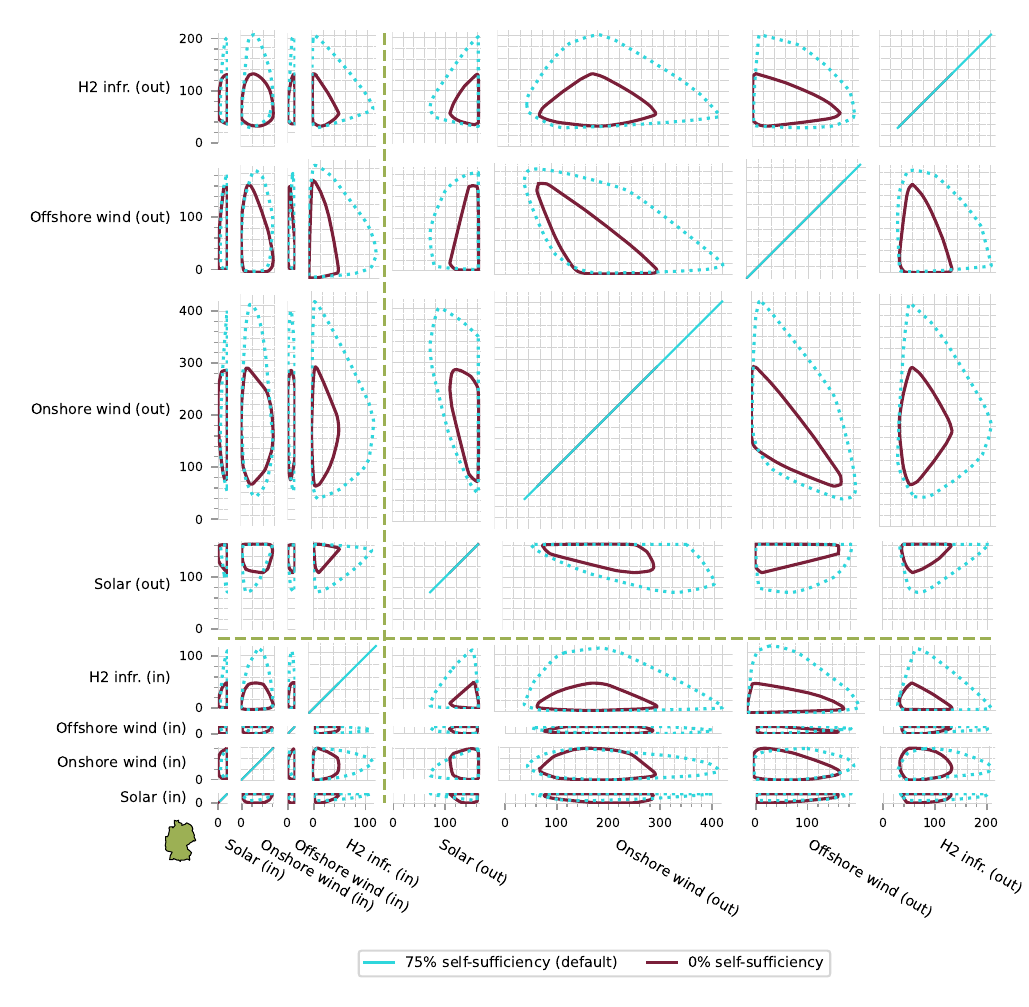}
  \caption[Near-optimal spaces for different self-sufficiency levels]{
    Projections of the intersection of near-optimal spaces for the German-focused model depending on whether the self-sufficiency constraint is enforced.
    Here, the ``(in)'' and ``(out)'' suffixes indicate that the respective dimensions denote investment inside and outside the given focus region (in this case, Germany).
    Note that with relaxing the self-sufficiency constraint the reduction of total system costs makes less money available for near-optimal solutions, shrinking the robust design space. The intersections are based on 150 iterations per near-optimal space (for the near-optimal spaces with 75\% self-sufficiency we only took the first 150 of the 450 iterations used in the study).}
  \label{fig:self-suff-sensitivity}
\end{figure*}

\clearpage

\begin{figure*}[h!]
  \centering
  \includegraphics{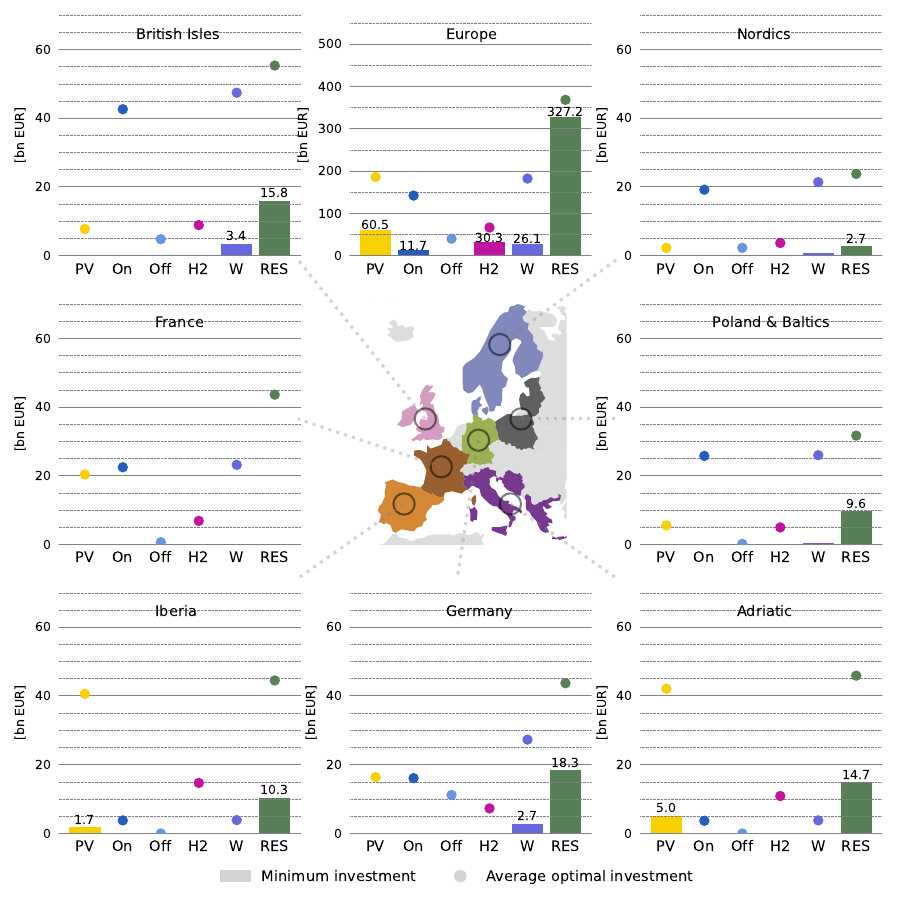}
  \caption[Version of Figure 2 without solar land-use scenarios]{
    Comparison of minimal regional and European robust investment (in bn EUR, annualised), excluding the restrictive land-use scenarios. This parallels Figure 2 in the main text, but restrictive solar land use scenarios are excluded as in the previous figure. Allowing for additional land-use for solar has a significant effect on minimum robust investment in wind power. For example, whereas Figure 2 in the main text shows a minimum investment of 12.3 and 165.8 bn EUR in wind power of any kind for the British Isles and all of Europe, respectively, the corresponding numbers in this figure (allowing for additional solar land use) are only 3.4 and 26.1 bn EUR, respectively. Minimum total robust investment in renewables (RES in this plot) is less affected.}
  \label{fig:min-investments-without-land-use}
\end{figure*}

\begin{figure*}[h!]
  \centering
  \includegraphics{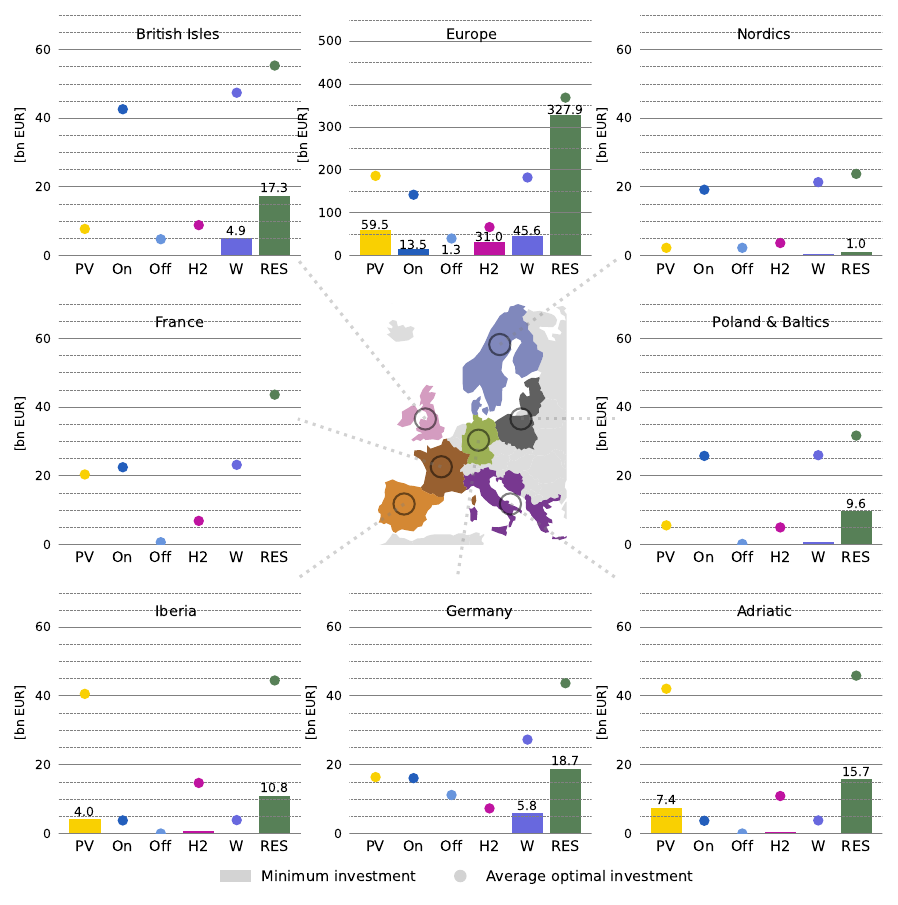}
  \caption[Version of Figure 2 with restrictive land-use for wind instead of solar]{Comparison of minimal regional and European robust investment (in bn EUR, annualised), excluding the restrictive land-use scenarios. This parallels Figure 2 in the main text, but instead of including scenarios where solar land-use is restricted to a third of the baseline, wind power land-use is restricted to a third of the baseline instead. Similar to \cref{fig:min-investments-without-land-use}, minimum investments in wind power especially are significant reduced as solar can be used to greater degree to compensate low investment in wind power.}
  \label{fig:min-investments-without-solar-land-use}
\end{figure*}

\begin{figure*}[h!]
  \centering
  \includegraphics{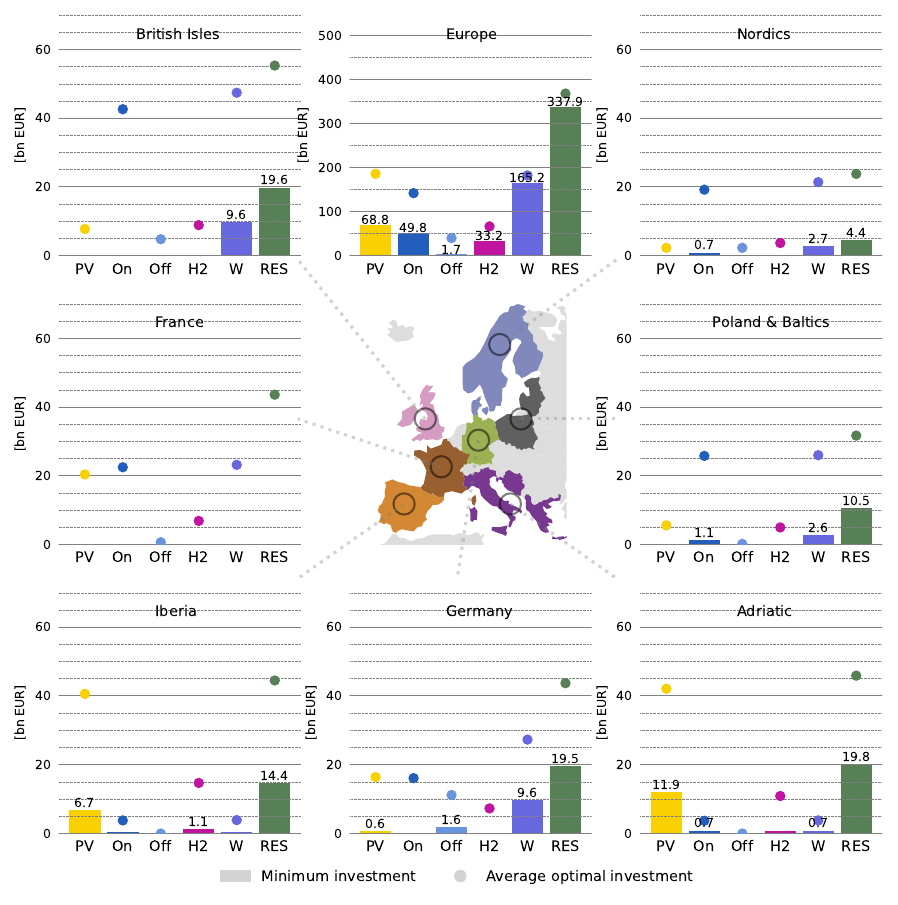}
  \caption[Version of Figure 2 with restrictive land-use for both wind and solar]{Comparison of minimal regional and European robust investment (in bn EUR, annualised), excluding the restrictive land-use scenarios. This parallels Figure 2 in the main text, but in addition to the scenarios limiting solar land-use by a third, we also include scenarios that limit wind power land-use by a third. Minimum investments are slightly higher than in Figure 2, but broadly similar. The most significant difference is a much higher minimum investment in solar in the Iberian peninsula.}
  \label{fig:min-investments-with-wind-land-use}
\end{figure*}

\clearpage

\begin{figure}[h!]
  \centering
  \includegraphics{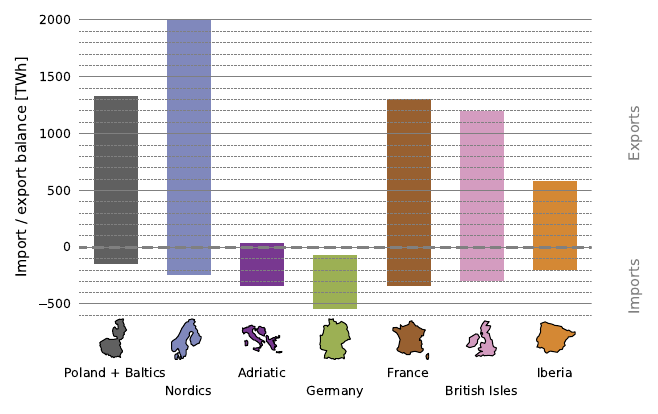}
  \caption[Robust ranges of net annual energy balance per focus region]{
    Robust ranges of net annual energy balance per focus region.
    Specifically, the bars indicate the minimum and maximum energy trade balances observed across a large sample of robust solutions.
    All regions can be net importers (while adhering to the 75\% self-sufficiency) and most can also be net exporters to various degrees.
    While the lower ends of the ranges essentially reflect the imposed constraint of 75\% self-sufficiency, the upper ends are determined by a combination of competitive renewable generation potential and access to export destinations.
    The absolute values of maximum possible exports may change with the system cost slack level (here 5\%) where export potential is limited by profitability and not land-use.
  }
  \label{fig:export-potential}
\end{figure}

\begin{mynote}[h!]
  \caption[Robust export ranges]{\textbf{Robust export ranges.}
  \Cref{fig:export-potential} shows robust ranges of exports/imports observed for the 7 different focus regions.
  To compute export potential, we need a representative sample of full system designs (including operations) laying inside $\mathcal{I}_\varepsilon^r$ for each region $r$ (Methods).
  This is because the geometric shape of $\mathcal{I}_\varepsilon^r$, which we compute as an intersection of near-optimal spaces, only contains information about total investment in solar, onshore wind, offshore wind and hydrogen (inside and outside $r$), but no total export figures.
  Thus, for each $r$, we sample 300 points in $\mathcal{I}_\varepsilon^r$, and run model optimisations where total solar, onshore wind, offshore wind and hydrogen investment are fixed to the coordinates of the 300 points.
  We can then calculate total imports / exports for each model and region; the results are shown in \cref{fig:export-potential}.
  The 300 points are sampled randomly on the boundary of $\mathcal{I}_\varepsilon^r$, except the first 24 which consist of those points which min- and maximise each individual of the 8 key dimensions (16 total) and those that min- and maximise the system-wide solar, onshore wind, offshore wind and hydrogen dimensions (8 total).}
\end{mynote}

\clearpage

\begin{figure*}[h!]
  \centering
  \includegraphics{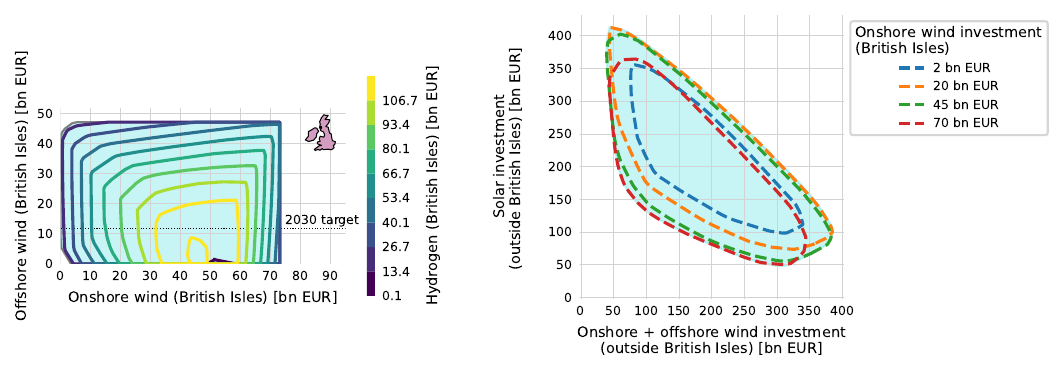}
  \caption[Version of Figure 3 without solar land-use scenarios]{Internal dynamics between wind power and hydrogen on the British Isles and the effects of British wind power on continental renewable investment, excluding the restrictive land-use scenarios. This parallels Figure 3 in the main text, but restrictive solar land use scenarios are excluded as in the previous figure. Note that lower amounts of British wind investment can be sustained without the tighter solar land-use restriction (compare the lower left of the left subplot to Figure 3 in the main text). Even with more solar allowed, the right subplot shows that minimum wind investment in the rest of Europe is still significantly affected by British wind investment.}
  \label{fig:combined-UK-regional-vs-continental-without-land-use}
\end{figure*}

\begin{figure*}[h!]
  \centering
  \includegraphics{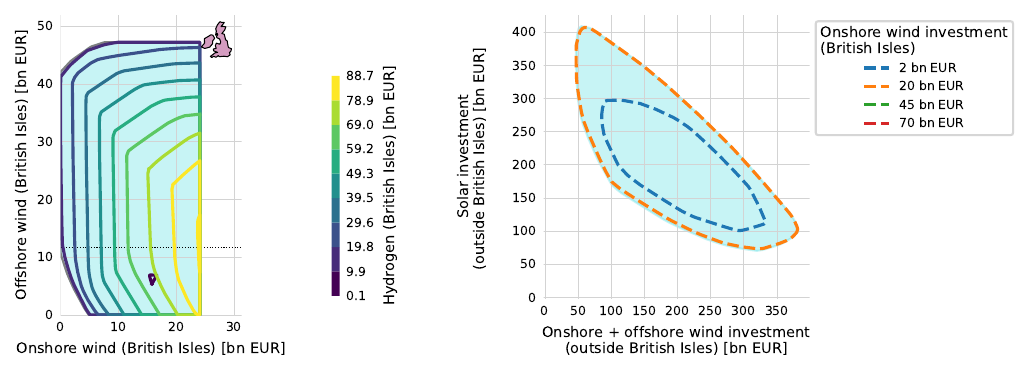}
  \caption[Version of Figure 3 with restrictive land-use for wind instead of solar]{Internal dynamics between wind power and hydrogen on the British Isles and the effects of British wind power on continental renewable investment, excluding the restrictive land-use scenarios. This parallels Figure 3 in the main text, but instead of including scenarios where solar land-use is restricted to a third of the baseline, wind power land-use is restricted to a third of the baseline instead.}
  \label{fig:combined-UK-regional-vs-continental-without-solar-land-use}
\end{figure*}

\begin{figure*}[h!]
  \centering
  \includegraphics{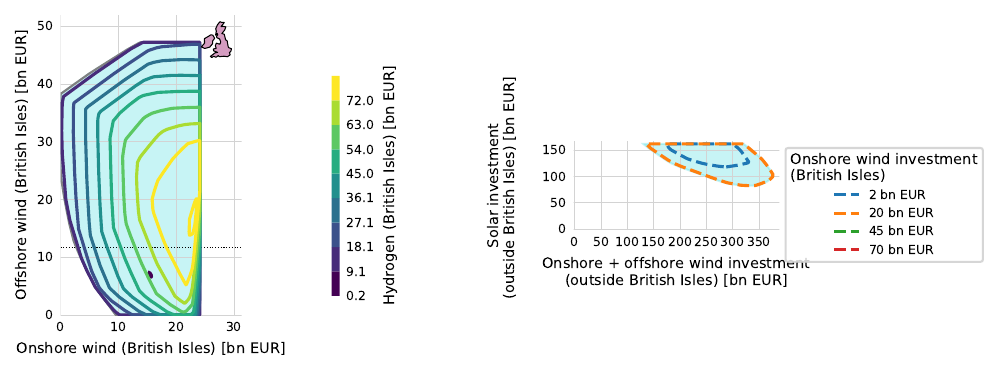}
  \caption[Version of Figure 3 with restrictive land-use for both wind and solar]{Internal dynamics between wind power and hydrogen on the British Isles and the effects of British wind power on continental renewable investment, excluding the restrictive land-use scenarios. This parallels Figure 3 in the main text, but in addition to the scenarios limiting solar land-use by a third, we also include scenarios that limit wind power land-use by a third.}
  \label{fig:combined-UK-regional-vs-continental-with-wind-land-use}
\end{figure*}

\clearpage

\begin{figure*}[h!]
  \centering
  \includegraphics{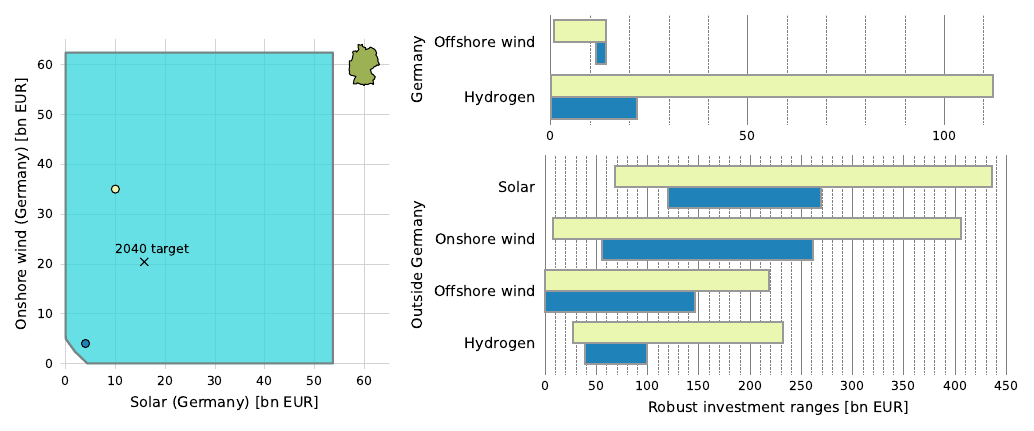}
  \caption[Version of Figure 4 without solar land-use scenarios]{Investment flexibility inside and outside Germany for two particular combinations of solar and onshore wind investment levels. This parallels Figure 4 in the main text, but restrictive solar land use scenarios are excluded as in the previous figure.}
  \label{fig:DE-focus-two-points-without-land-use}
\end{figure*}

\begin{figure*}[h!]
  \centering
  \includegraphics{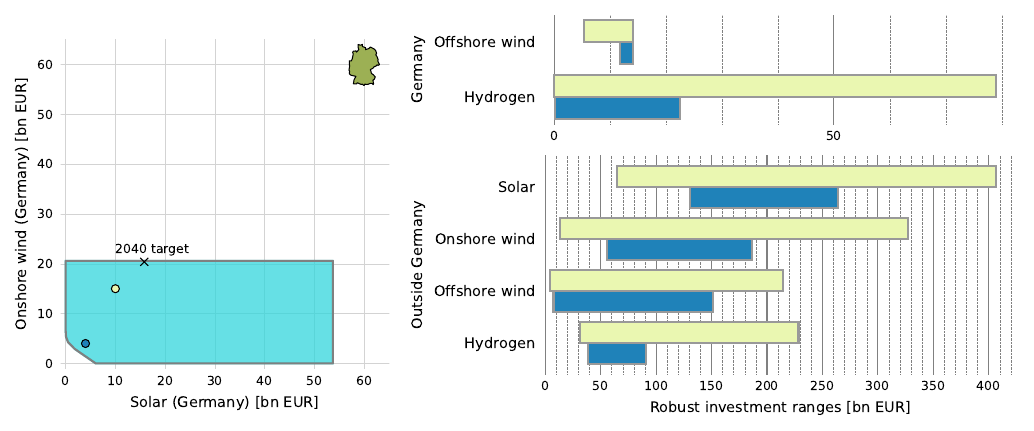}
  \caption[Version of Figure 4 with restrictive land-use for wind instead of solar]{Investment flexibility inside and outside Germany for two particular combinations of solar and onshore wind investment levels. This parallels Figure 4 in the main text, but instead of including scenarios where solar land-use is restricted to a third of the baseline, wind power land-use is restricted to a third of the baseline instead.}
  \label{fig:DE-focus-two-points-without-solar-land-use}
\end{figure*}

\begin{figure*}[h!]
  \centering
  \includegraphics{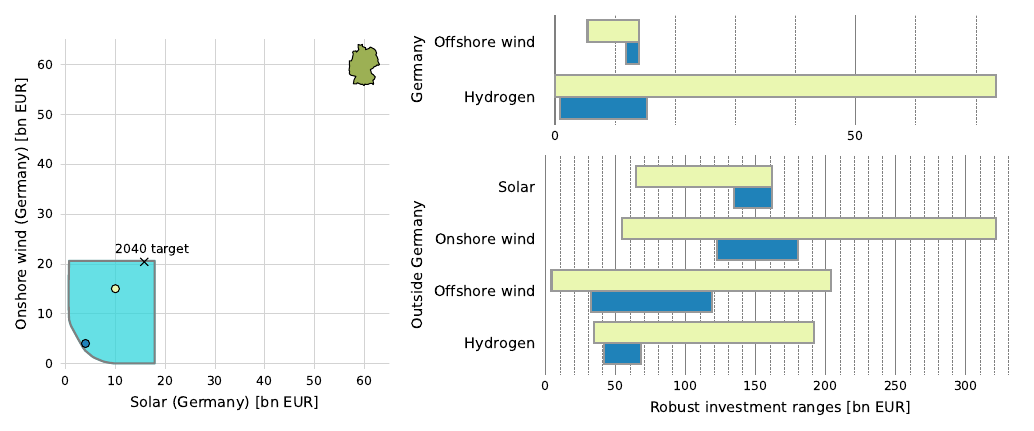}
  \caption[Version of Figure 4 with restrictive land-use for both wind and solar]{Investment flexibility inside and outside Germany for two particular combinations of solar and onshore wind investment levels. This parallels Figure 4 in the main text, but in addition to the scenarios limiting solar land-use by a third, we also include scenarios that limit wind power land-use by a third.}
  \label{fig:DE-focus-two-points-with-wind-land-use}
\end{figure*}

\clearpage

\begin{figure*}[h!]
  \centering
  \includegraphics{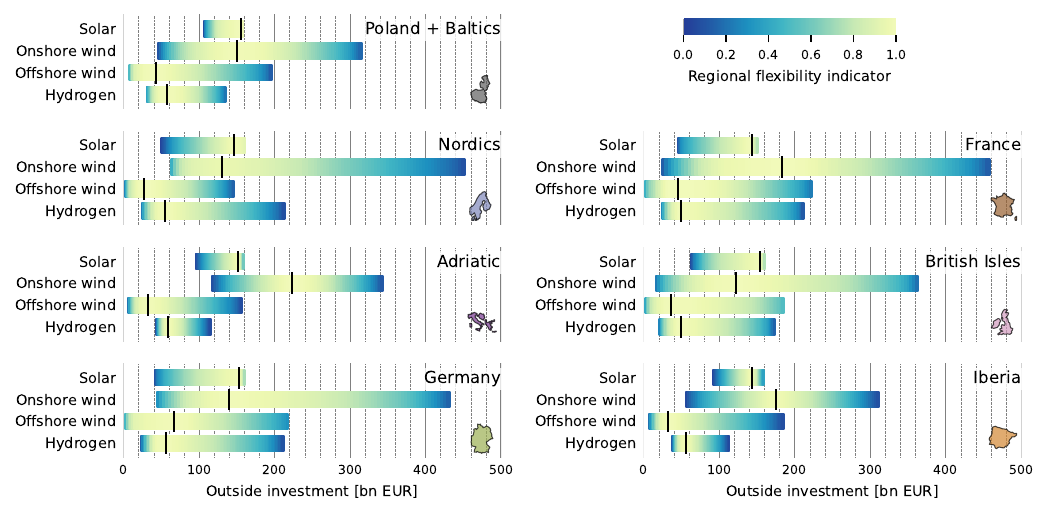}
  \caption[Effects of continental investment decisions on the design flexibility of the focus regions]{Effects of continental investment decisions on the design flexibility of the focus regions.
    The calculation of the flexibility indicator (where 1 is maximally flexible) that is plotted is explained in the Methods.
    The levels of continental investment leading to maximal regional flexibility are marked with black vertical bars.
    For instance, European investment in onshore wind has a positive effect on design flexibility in Poland and the Baltic states from 40 and up to about 150 bn EUR, with only a marginal decrease in flexibility upon further investment.
    Related to Figure 5.
  }
  \label{fig:continental-investment-impact}
\end{figure*}

\begin{figure*}[h!]
  \centering
  \includegraphics{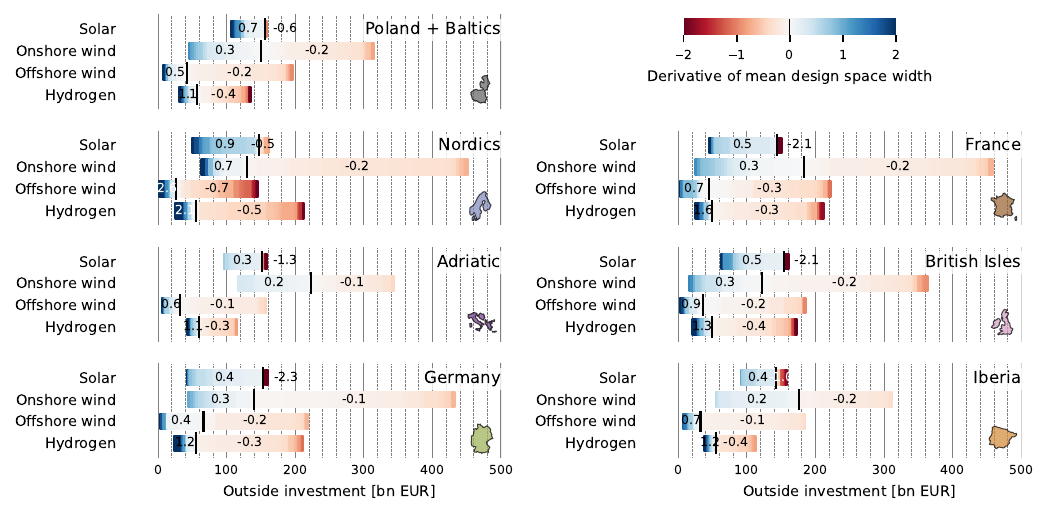}
  \caption[More detailed version of \cref{fig:continental-investment-impact} using derivates]{Derivative of mean width of focus region design space with respect to changes in investment outside the focus region.
    The levels of outside investment leading to maximal regional flexibility (where the derivative is zero) are marked with black vertical bars.
    On either side, the derivative of mean width of the robust design space of the given focus region, with respect to outside investment in solar, onshore- and offshore wind and hydrogen, is indicated using a colour scale.
    A positive derivative indicates that additional marginal investment in a given technology (outside a focus region) enlarges the design space, thus the flexibility, of the focus region.
    For each region and technology, the mean derivatives up to maximum flexibility and beyond this point (which are positive and negative respectively) are annotated.
    Related to \cref{fig:continental-investment-impact}.
  }
  \label{fig:derivative-inside}
\end{figure*}

\begin{figure*}[h!]
  \centering
  \includegraphics{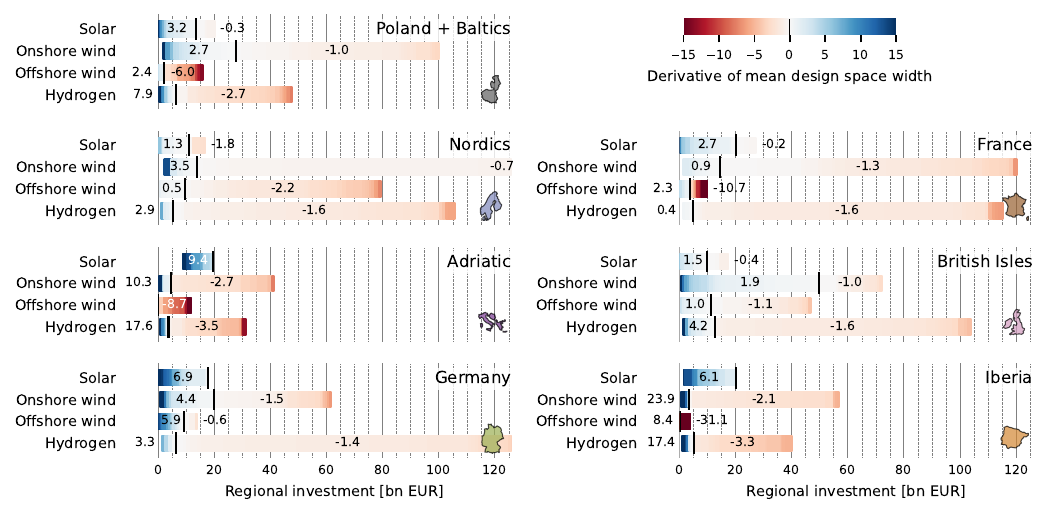}
  \caption[More detailed version of Figure 5 using derivates]{Derivative of mean width of the design space outside the focus region with respect to changes in investment inside the focus region.
    The levels of inside investment leading to maximal system-wide flexibility (where the derivative is zero) are marked with black vertical bars.
    On either side, the derivative of mean width of the robust design space outside the given focus region, with respect to inside investment in solar, onshore- and offshore wind and hydrogen, is indicated using a colour scale.
    A positive derivative indicates that additional marginal investment in a given technology (inside a focus region) enlarges the design space, thus the flexibility, outside the focus region.
    For each region and technology, the mean derivatives up to maximum flexibility and beyond this point (which are positive and negative respectively) are annotated.
    Related to Figure 5.
  }
  \label{fig:derivative-outside}
\end{figure*}

\begin{mynote}
  \caption[Impact of regional/European investment on the near-optimal space]{\textbf{Impact of regional/European investment on the near-optimal space.}
  In Figure 7 in the main text, we present the impact of regional investment on robust European investment through a flexibility indicator (defined in the Methods).
  The flexibility indicator in this case is based on the mean width of the space of robust designs (projected onto the 4 key dimensions of European investment), depending on (robust) investment levels for each regional key dimension.
  In other words, for a given inside investment level (say, a 10 bn EUR investment in onshore wind in the Nordics), we compute outside design flexibility as the mean of the differences between maximum and minimum outside investment in solar, onshore wind, offshore wind and hydrogen.
  The following plot shows the ``converse'': the impact of continental investment on regional design flexibility.
  The plot is analogous to Figure 7, but the roles of ``inside'' and ``outside'' have been swapped.

  Related to this, it is possible to identify at what (regional or continental) investment levels the largest gains or losses of design flexibility are realised:
  by taking the derivative of the mean width of the robust design space --- formally the derivative of the absolute (not relative as in Figure 5 and \cref{fig:continental-investment-impact}) flexibility indicator.
  By varying technology investment levels \emph{inside} a focus region as in \cref{fig:derivative-inside} we capture the increasing or decreasing mean width of the projection onto European investments \emph{outside} the focus region.
  These increases and decreases of flexibility are shown as derivatives in \cref{fig:derivative-inside} and for the converse (based on outside investment variations) in \cref{fig:derivative-outside}.}
\end{mynote}

\clearpage

\begin{figure*}[h!]
  \centering
  \includegraphics[width=\textwidth]{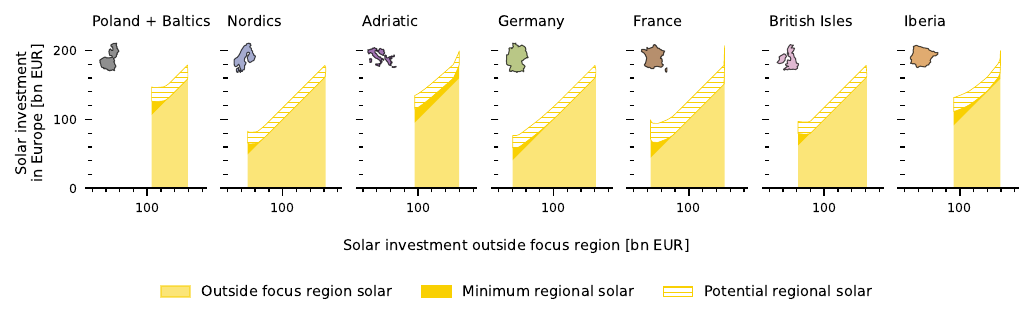}
  \caption[Ranges of regional solar investment as function of outside solar investment]{Minimum and maximum potential ranges of total solar power investment in different regions, as a function of solar power investment outside the respective regions.
    Note that each subplot represents separate results from differently focused models, hence the disagreement on maximum overall viable solar power investment. Related to Figure 6.}
  \label{fig:solar-ranges}
\end{figure*}

\begin{figure*}[h!]
  \centering
  \includegraphics[width=\textwidth]{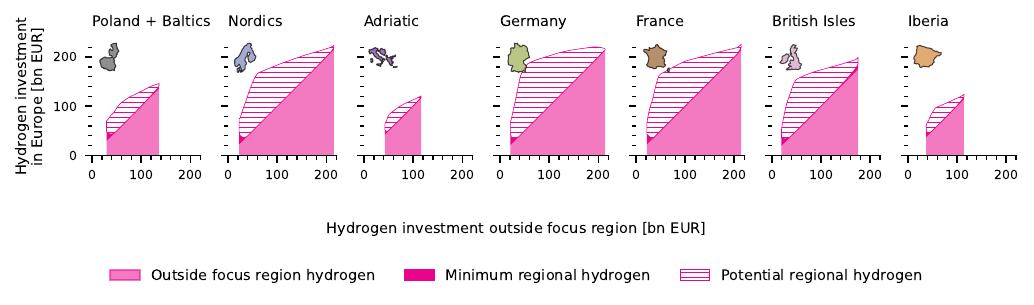}
  \caption[Ranges of regional hydrogen investment as function of outside solar investment]{Minimum and maximum potential ranges of total hydrogen infrastructure investment in different regions, as a function of hydrogen infrastructure investment outside the respective regions.
    Note that each subplot represents separate results from differently focused models, hence the disagreement on maximum overall viable hydrogen infrastructure investment. Related to Figure 6.}
  \label{fig:hydrogen-ranges}
\end{figure*}

\clearpage

\begin{figure*}[h!]
  \centering
  \includegraphics[width=\textwidth]{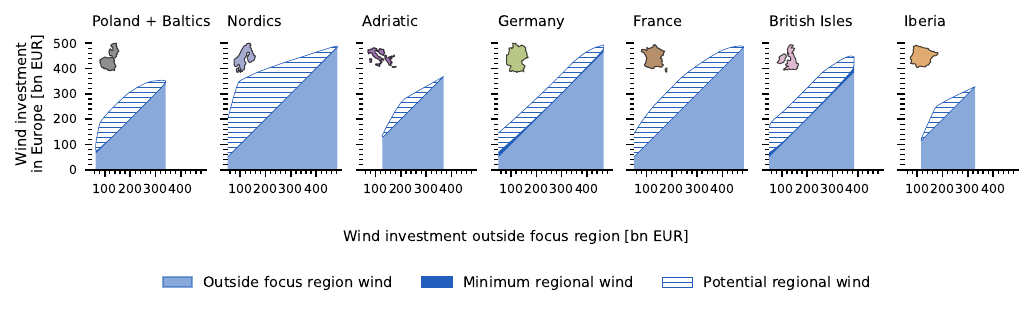}
  \caption[Version of Figure 6 without solar land-use scenarios]{Minimum and maximum potential ranges of total wind (onshore + offshore) investment in different regions, as a function of total wind investment outside the respective regions. This figure parallels Figure 6 in the main text, but restrictive solar land use scenarios are excluded as in the previous figure. In line with \cref{fig:min-investments-without-land-use}, minimum regional wind investment is significantly reduced compared to Figure 6.}
  \label{fig:wind-ranges-without-land-use}
\end{figure*}

\begin{figure*}[h!]
  \centering
  \includegraphics[width=\textwidth]{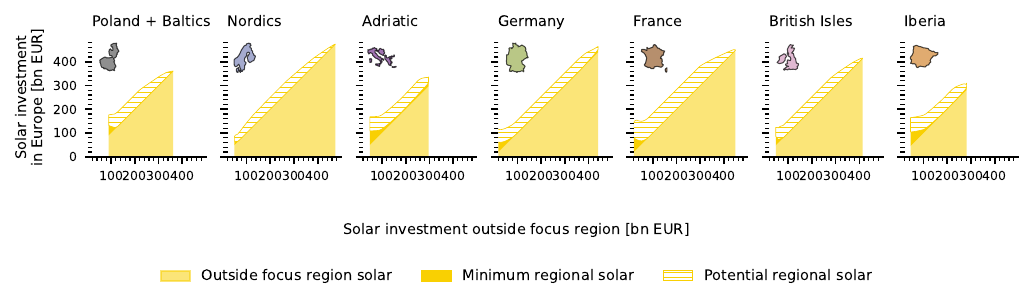}
  \caption[Version of \cref{fig:solar-ranges} without solar land-use scenarios]{Minimum and maximum potential ranges of total solar power investment in different regions, as a function of solar power investment outside the respective regions.
    This figure parallels \cref{fig:solar-ranges}, but restrictive solar land use scenarios are excluded as in the previous figure. Compared to \cref{fig:solar-ranges}, we see that regional solar investment can be out-competed by very high outside solar investment.
  }
  \label{fig:solar-ranges-without-land-use}
\end{figure*}

\begin{figure*}[h!]
  \centering
  \includegraphics[width=\textwidth]{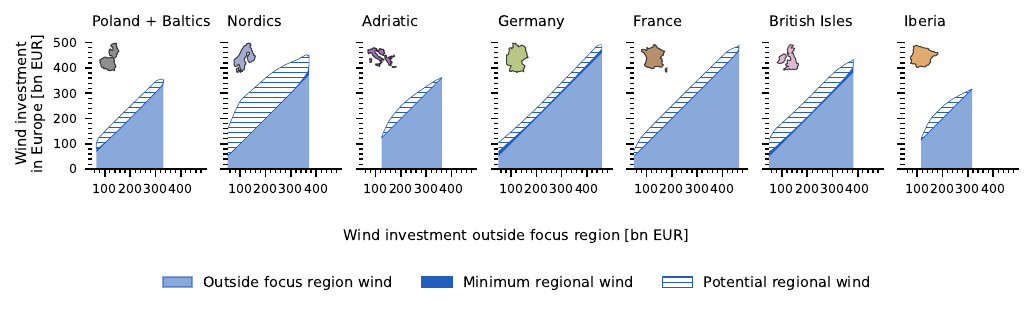}
  \caption[Version of Figure 6 with restrictive land-use for wind instead of solar]{Minimum and maximum potential ranges of total wind (onshore + offshore) investment in different regions, as a function of total wind investment outside the respective regions. This figure parallels Figure 6 in the main text, but instead of including scenarios where solar land-use is restricted to a third of the baseline, wind power land-use is restricted to a third of the baseline instead.}
  \label{fig:wind-ranges-without-solar-land-use}
\end{figure*}

\begin{figure*}[h!]
  \centering
  \includegraphics[width=\textwidth]{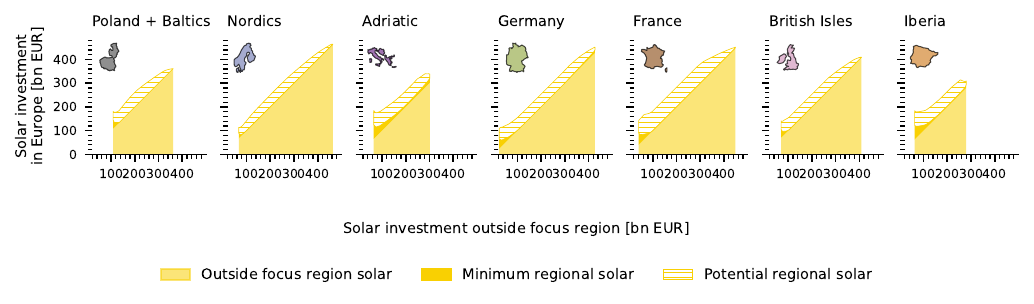}
  \caption[Version of \cref{fig:solar-ranges} with restrictive land-use for wind instead of solar]{Minimum and maximum potential ranges of total solar power investment in different regions, as a function of solar power investment outside the respective regions.
    This figure parallels \cref{fig:solar-ranges}, but instead of including scenarios where solar land-use is restricted to a third of the baseline, wind power land-use is restricted to a third of the baseline instead.
  }
  \label{fig:solar-ranges-without-solar-land-use}
\end{figure*}

\begin{figure*}[h!]
  \centering
  \includegraphics[width=\textwidth]{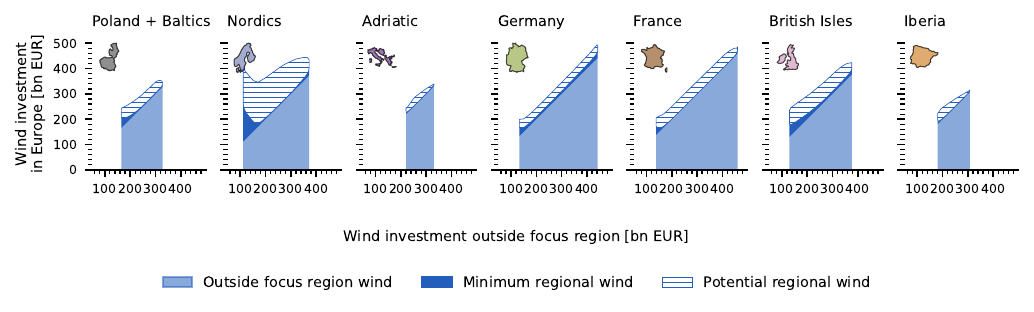}
  \caption[Version of Figure 6 with restrictive land-use for both wind and solar]{Minimum and maximum potential ranges of total wind (onshore + offshore) investment in different regions, as a function of total wind investment outside the respective regions. This figure parallels Figure 6 in the main text, but in addition to the scenarios limiting solar land-use by a third, we also include scenarios that limit wind power land-use by a third.}
  \label{fig:wind-ranges-with-wind-land-use}
\end{figure*}

\begin{figure*}[h!]
  \centering
  \includegraphics[width=\textwidth]{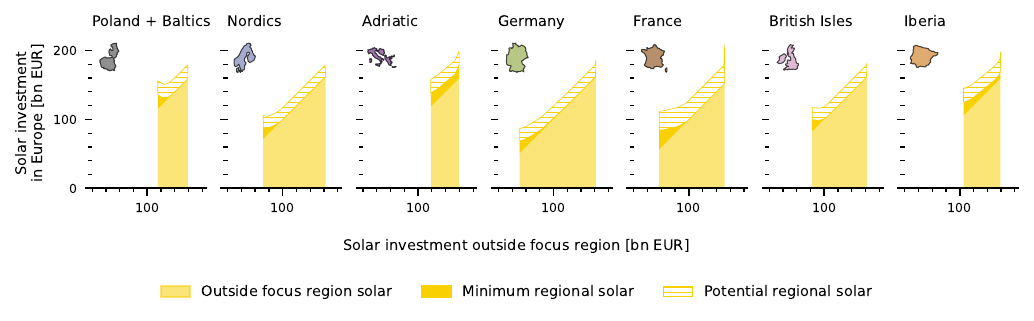}
  \caption[Version of \cref{fig:solar-ranges} with restrictive land-use for both wind and solar]{Minimum and maximum potential ranges of total solar power investment in different regions, as a function of solar power investment outside the respective regions.
    This figure parallels \cref{fig:solar-ranges}, but in addition to the scenarios limiting solar land-use by a third, we also include scenarios that limit wind power land-use by a third.}
  \label{fig:solar-ranges-with-wind-land-use}
\end{figure*}

\clearpage

\begin{figure*}[h!]
  \centering
  \includegraphics{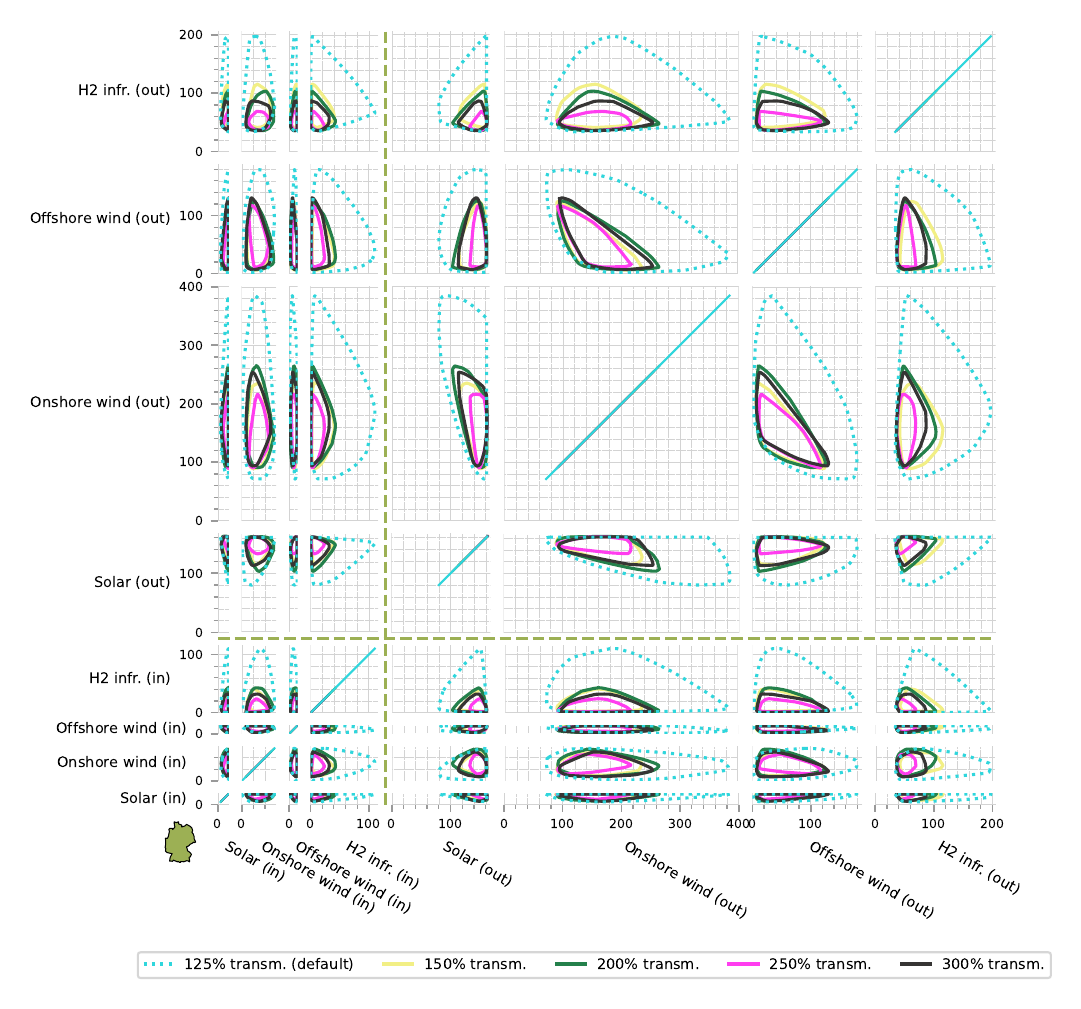}
  \caption[Impact of transmission expansion limitations on the near-optimal space]{
    Projections of the intersection of near-optimal spaces for the German-focused model depending on the constraint of transmission expansion.
    Here, the ``(in)'' and ``(out)'' suffixes indicate that the respective dimensions denote investment inside and outside the given focus region (in this case, Germany).
    Note that for higher allowed transmission expansion the reduction of total system costs makes less money available for near-optimal solutions, shrinking the robust design space. All five intersections are based on 140 iterations per near-optimal space (for the near-optimal spaces with 125\% transmission expansion we only took the first 140 of the 450 iterations used in the study).}
  \label{fig:transmission-sensitivity}
\end{figure*}

\begin{mynote}
  \caption[Sensitivity of transmission expansion limitations]{\textbf{Sensitivity of transmission expansion limitations.}
  The main results in the present study are based on optimisations in which we limit the transmission volume to 125\% of the currently existing transmission in Europe. Single transmission lines or links can still be expanded freely as long as the total volume of transmission in the system adheres to the 125\% constraint. We choose this conservative constraint, as most benefits in the power system are reaped at this level (Fig. 6 in H\"{o}rsch \& Brown \cite{horsch-brown-2017}); still, hydrogen transportation can still lead to significant energy transportation across the system.

  We have conducted a sensitivity analysis for one of the focus regions (Germany) to investigate the impact of different transmission constraints (150\%, 200\%, 300\%, 400\% compared to the default 125\%) on the robust design spaces and key system values (\cref{tab:transmission}). Note that in \cref{fig:transmission-sensitivity} the robust design spaces are smaller for higher transmission levels: the total system costs are lower when transmission can be expanded more, which reduces the upper bound on near-optimality from the definition of the design spaces ($\mathcal{F}_{\varepsilon}^{r,s}$ in Methods). The reduction in available capital has a larger impact on the robust solution space than being able to expand transmission more.}
\end{mynote}

\begin{table}[h]
  \centering
  \begin{tabular}{p{4.5cm} ccccc}

    \toprule
    Optimal investment (bn EUR) & 125\% trans. (default) & 150\% trans. & 200\% trans. & 300\% trans. & 400\% trans. \\
    \midrule
    Solar & 182.7 & 178.6 & 172.6   & 167.6  &  167.6  \\
    Onshore wind & 149.9 & 150.3 &  150.6  &  153.0 &  153.0 \\
    Offshore wind & 28.5 & 28.9 & 30.1   &  29.7 &  29.7  \\
    Hydrogen infrastructure & 67.8 & 63.1 &  56.8  &  54.6  & 54.6 \\
    Wind investment & 178.5 & 179.2 &  180.6  &  182.8  & 182.8 \\
    Renewable generation & 361.1 & 357.8 &  353.5  &  350.3  & 350.3 \\
    Total system costs & 779.6 & 761 & 754.5  &  752.9  & 752.9 \\
    \bottomrule
  \end{tabular}
  \caption[Sensitivity of cost-optimal system design to transmission expansion limits]{Optimal investment in Europe (average across the 12 scenarios) depending on the upper limit of transmission expansion: 125\%, 150\%, 200\%, 300\%, 400\% compared to current levels. The sensitivity analysis is conducted with Germany as the focus region. European investment is therefore the sum of inside (German) and outside (rest of the system) investments in solar, onshore and offshore wind and hydrogen infrastructure.}
  \label{tab:transmission}
\end{table}

\clearpage

\begin{figure*}[h]
  \centering
  \includegraphics{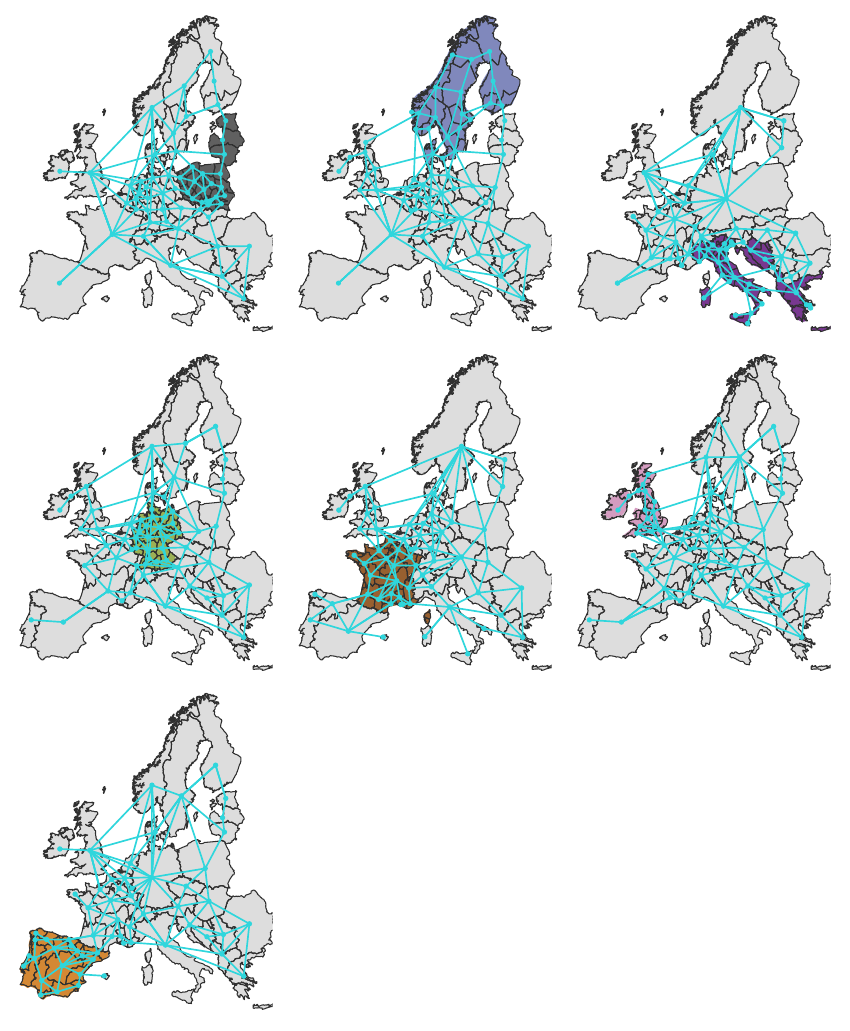}
  \caption[Model network topology for different focus regions]{Network topology for the various focus models. The plots only illustration spatial resolution: bus and line sizes do not carry any significance in this plot.
  As described in the Methods, we use different spatial layouts of the same model for our different focus regions.
This is in order to better represent spatial variability and transmission bottlenecks in the respective focus regions, while reducing the overall spatial resolution of the model for computation reasons.

For each focus region, we ``distribute'' 60 nodes of spatial resolution by giving a certain number of nodes to each of the focus region itself, the countries bordering the focus region, and the remaining countries.
The exact numbers of nodes allocated to each of these three categories depends on the focus region, and are listed in the Methods section of the main text.
Note that there may sometimes be fewer nodes than countries in the modelling area not bordering on the focus region; in this case whole countries may be clustered together (e.g. Germany, Czechia and Poland in the Adriatic-focused model).}
  \label{fig:regions}
\end{figure*}

\begin{figure*}[h!]
  \centering
  \includegraphics{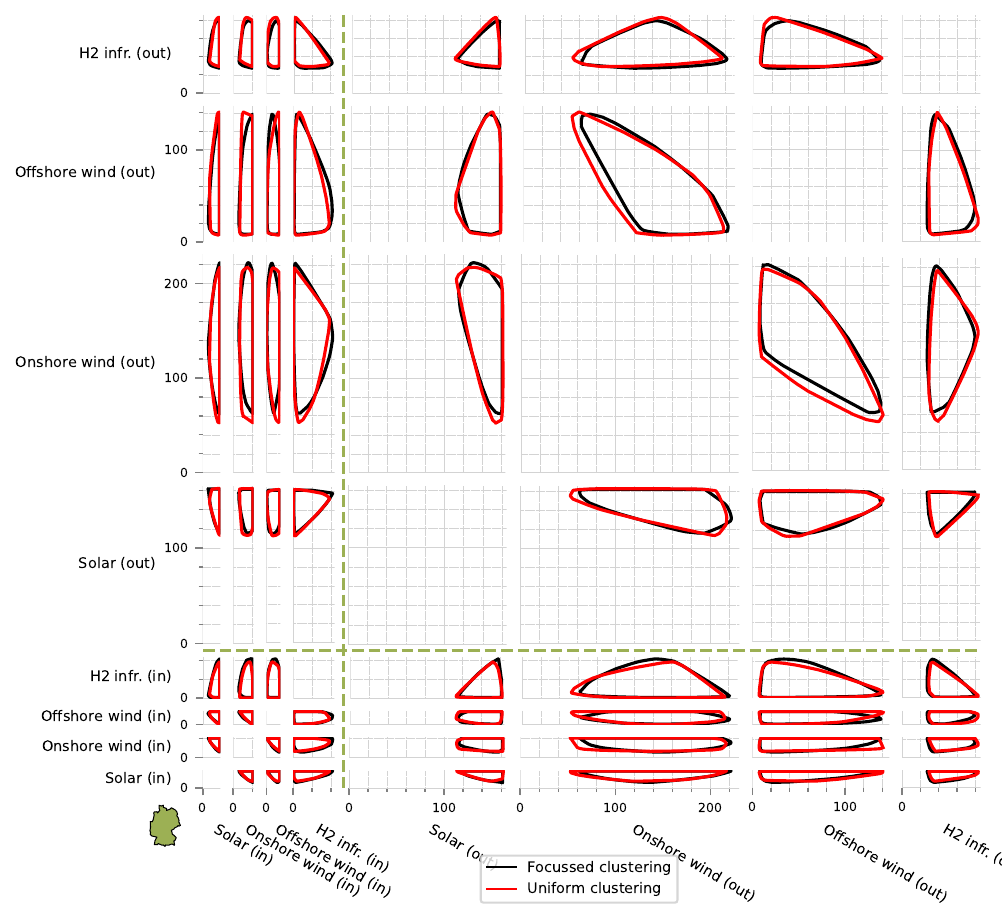}
  \caption[Comparison of near-optimal spaces with focussed and uniform clustering]{
    A comparison of the intersections of near-optimal spaces obtained using a model with focussed clustering (as shows in \cref{fig:regions}), and one with default (``uniform'') clustering. In this example, we show a German-focussed model with reduced (50-segment) temporal resolution.
    Here, the ``(in)'' and ``(out)'' suffixes indicate that the respective dimensions denote investment inside and outside the given focus region (in this case, Germany).
  }
  \label{fig:focussed-vs-uniform}
\end{figure*}

\begin{figure*}[h!]
  \centering
  \includegraphics{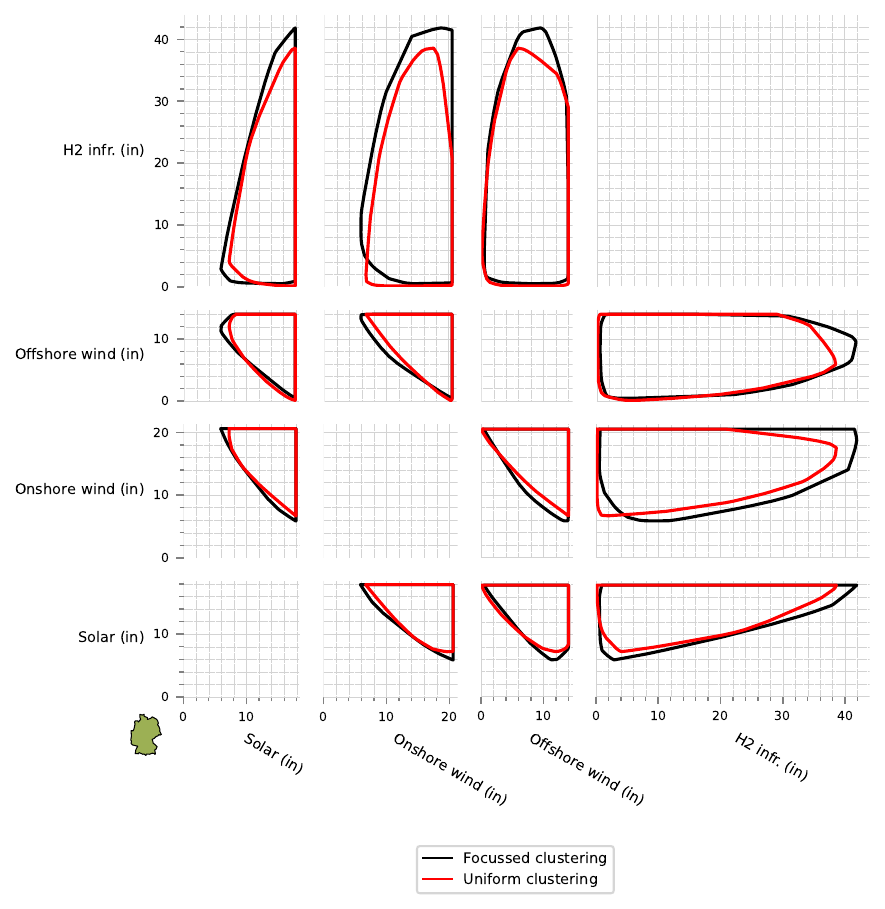}
  \caption[Comparison of intersection of near-optimal spaces with focussed and uniform clustering]{A comparison of the intersections of near-optimal spaces obtained using a model with focussed clustering (as shows in \cref{fig:regions}), and one with default (``uniform'') clustering. Here, only the ``inside'' dimensions are shown for Germany; the lower left quarter of \cref{fig:focussed-vs-uniform}.}
  \label{fig:focussed-vs-uniform-regional}
\end{figure*}

\clearpage

\begin{figure*}[h!]
  \centering
  \includegraphics{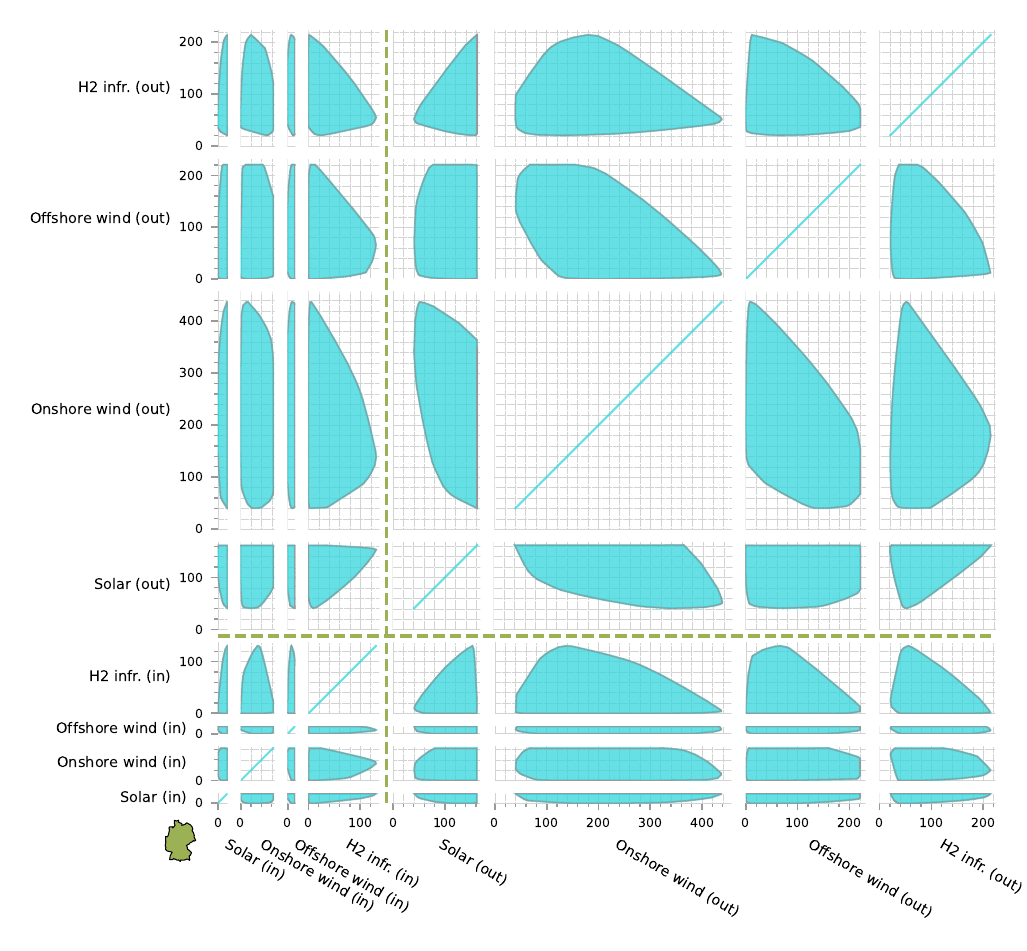}
  \caption[Projections of the intersection of near-optimal spaces for the German-focused model]{
    Projections of the intersection of near-optimal spaces for the German-focused model.
    We show projections onto each pair of the 8 dimensions $\{\text{Solar}, \text{Onshore wind}, \text{Offshore wind}, \text{Hydrogen}\} \times \{\text{inside}, \text{outside}\}$ of the robust design space for the German-focused model.
    Here, the ``(in)'' and ``(out)'' suffixes indicate that the respective dimensions denote investment inside and outside Germany.
    Recall that this robust design space is the intersection of near-optimal spaces arising from 12 different scenarios.
  }
  \label{fig:near-opt-grid}
\end{figure*}

\begin{figure*}[h!]
  \centering
  \includegraphics{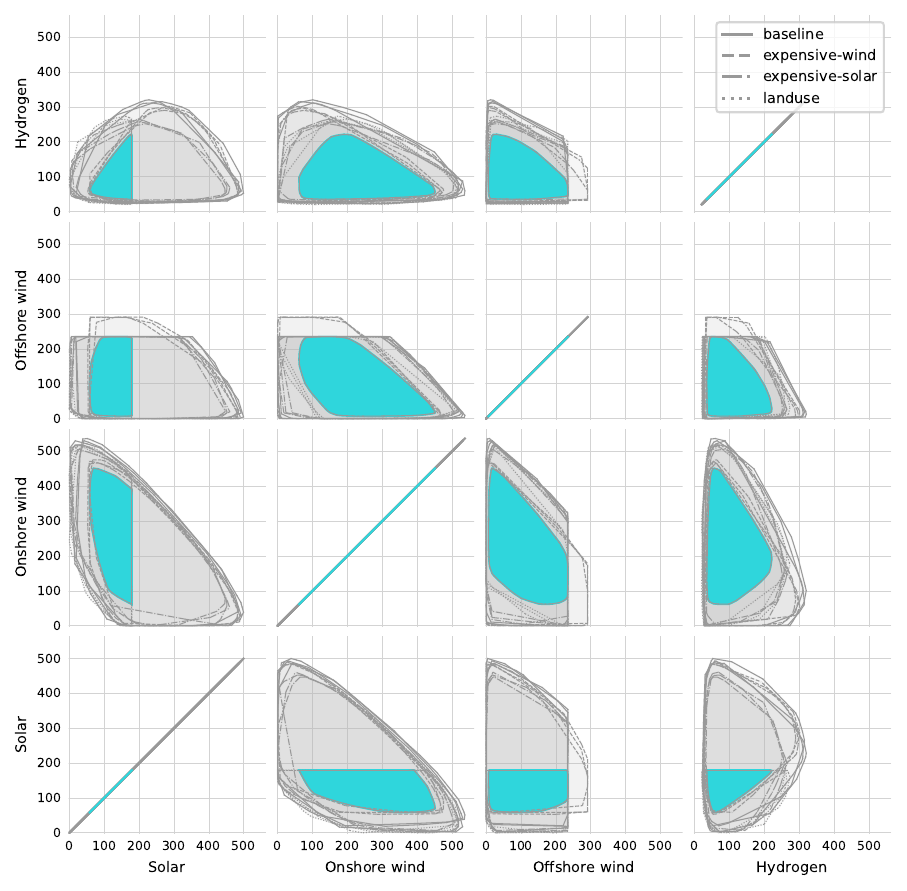}
  \caption[Projections of all near-optimal space for the German-focused model]{
    Projections onto four major dimensions of approximations of the near-optimal scenarios corresponding to 12 scenarios of the German-focused model (grey) as well as projections of the intersection of these scenarios (turquoise).
    Scenario categories are indicated with different line styles (see legend); for each category there are three different near-optimal spaces corresponding to the weather years 1985, 1987 and 2010.
    Each of the dimensions shows in this plot (solar, onshore wind, offshore wind, hydrogen) represent aggregations of the respective ``inside'' and ``outside'' dimensions.}
  \label{fig:near-opt-grid-europe}
\end{figure*}